# Cryohydrologic Flow and Pingo Formation Within Freezing Impact Melts in Occator Crater, Ceres

**A. R. Kubas[1], H. G. Sizemore[2], B. E. Schmidt[1]**

[1]Cornell University (122 Sciences Dr, Ithaca, NY 14850)
[2]Planetary Science Institute (1700 East Fort Lowell, Suite 106, Tucson, AZ)

Corresponding author: Alexia Kubas (ak2248@cornell.edu)

## Key Points

- Under most conditions, freezing impact melt can deliver sufficient water to Ceres' near surface to support pingo formation.
- Impact melt freezing creates sufficient stress to fracture the subsurface, offering additional pathways for groundwater flow.

## Abstract

Ceres is a volatile-rich dwarf planet in the asteroid belt and a potential relict ocean world. It exhibits surface features indicative of subsurface brine activity, particularly within Occator Crater. High-resolution imagery from NASA's Dawn Mission reveals landslides, gullies, carbonate faculae, and mound-like features resembling terrestrial pingos on the floor of Occator Crater. Some of these landforms have been hypothesized to form from water, either from the ascent of deep crustal brines or from the freezeout of impact melt. Here, we test the hypothesis that pingo-like features can be formed via the freezing of shallow impact melts. We simulate the heat transfer and groundwater flow dynamics of a freezing impact melt deposit on the floor of Occator Crater. We test a comprehensive parameter space of regolith physical and material properties relevant to the cerean environment. We find that impact melt freezing produces pressure gradients sufficient to drive hydrologic flow within the melt matrix and could transport fluids to the near surface through hydraulic fracturing. For most of the parameter space tested, water volumes delivered to pingo candidates on the surface meet or exceed those likely required for pingo formation. Analogous hydrologic processes may be important in the shallow subsurface of other regions of Occator and Ceres at large as well as other ice-rich planetary bodies.

## Plain Language Summary

We test the hypothesis that hills located in impact melt deposits in Occator Crater, Ceres could have formed from a process similar to the one that forms terrestrial pingos. Pingos are hills made of permafrost and ice that form by the flow and freezing of groundwater near Earth's poles. If pingos exist on Ceres, they may be sources of abundant ice in the shallow subsurface and would be compelling targets in future space exploration missions. We find that, under conditions relevant to Ceres, enough groundwater is delivered by the impact melt freezing process to form pingos on the surface.

## 1 Introduction

Dwarf planet Ceres is ice-rich and may have a liquid brine layer in its deep interior (Raymond et al. 2020). Water ice is likely abundant in the upper ~10 km of Ceres' crust, as evidenced by geomorphology (e.g., Schmidt et al. 2017, Sizemore et al. 2019), geophysical inversions from Dawn gravity and shape data (Ermakov et al. 2017; Park et al. 2020), crater scaling (Hiesinger et al. 2016), and neutron and gamma-ray spectroscopy (Prettyman et al. 2017). In 2018, NASA's Dawn extended mission conducted several close orbital passes of the dwarf planet, allowing for high-resolution (3–20 m/pix) imaging that revealed evidence of a volatile-rich interior and surface. Modeling of Ceres' thermal evolution informed by Dawn datasets suggests that Ceres is a relict ocean world, which may still have localized brine reservoirs at depth (> ~40 km). Further, deep brine may periodically find connective pathways to the surface following impact or tectonic events (Castillo-Rogez & McCord 2010; Quick et al. 2019; Neveu and Desch 2015; Castillo-Rogez et al., 2018; Ruesch et al. 2017; Stein et al. 2017; Zolotov 2017; Raymond et al. 2020).

Occator Crater is a young, 92 km wide complex crater on Ceres, notable for its bright carbonate deposits on the crater floor (i.e., faculae). These faculae have been hypothesized to form from the cratering process itself or later extrusive cryovolcanic activity from a deep brine reservoir beneath Occator Crater (Quick et al. 2019; Ruesch et al. 2019; Stein et al. 2019; Scully et al. 2020). Many studies have investigated the thermodynamical evolution of Occator Crater and potential sources of its faculae (Quick et al. 2019, Ruesch et al. 2019, Hesse & Castillo-Rogez 2019; Bowling et al. 2019; Raymond et al. 2020; Stein et al. 2019; Scully et al. 2020). Morphologic studies of crater facies also suggest that ice and other volatiles are a major constituent of the shallow subsurface at Occator, similar to Ceres' bulk composition (Schmidt et al 2017, Sizemore et al., 2017; Schmidt et al. 2020; Scully et al. 2020, Schenk et al. 2020). Framing Camera (FC) images from Dawn's Low Altitude Mapping Orbit (LAMO) and extended mission revealed numerous small mounds on the floor of Occator Crater morphologically reminiscent of terrestrial pingos (Schmidt et al. 2020; Hughson et al. 2022).

Pingos are ice-cored hills that form in periglacial landscapes on Earth via the flow and subsequent freezing of pressurized groundwater. Terrestrial pingos typically form in valleys between hillslopes ("hydraulic" pingos) or in the centers of drained thermokarst lakes ("hydrostatic" pingos) (Mackay 1979). Vertical growth occurs seasonally and on longer timescales, as groundwater is injected to the pingo system through freeze-thaw processes. Mature terrestrial pingos develop heavily fractured summits, due to rupture driven by internal ice core development. Some hills on the floor of Occator Crater exhibit strikingly similar morphologies to terrestrial pingos. Moreover, their setting in volatile-rich, frozen impact melts suggests a potential genetic similarity (Figure 1; Hughson et al. 2022). With an axial tilt of 4º (Schorghofer et al. 2016), Ceres lacks the seasonal, interannual, and climatological freeze-thaw cycles that dominate terrestrial pingo evolution (see section 3.3). However, topographic analysis indicates that conditions for the development of both hydraulic and hydrostatic pingos are plausible on the floor of Occator (Schmidt et al. 2020). Although Ceres does not experience seasonal freeze-thaw processes, the freezing of an impact melt slurry mimics seasonal to climatic timescale freezing of terrestrial permafrost and is broadly consistent with planetary periglacial environments as understood in the martian context (e.g., Mellon et al., 2009). If pingos exist on Ceres, they are reservoirs of abundant near surface ice and would make excellent candidates for future missions focused on in situ resource utilization. The preservation of microbes and biomarkers in massive ice on Earth (e.g. Delpech et al. 2024) and the potential for films of unfrozen water in the subsurface makes pingos prominent, spacecraft-accessible astrobiology targets, as well.

Ceres' ice-rich crust shares several characteristics with periglacial landscapes, including a shallow ice table similar to that observed on Mars and in the Antarctic Dry Valleys (Prettyman et al. 2017; Hughson et al. 2018; Hughson et al. 2019; Sizemore et al., 2019, Chilton et al 2019; Prettyman et al. 2021), abundant morphological evidence for viscous relaxation and creep (Bland et al., 2018; Sori et al., 2018; Sizemore et al., 2019), localized morphologic evidence of ice-loss via sublimation (Sizemore et al. 2017; Sizemore et al. 2019), and numerous small hills similar to pingos inside Occator and Urvara Craters (Schenk et al. 2019; Sizemore et al., 2019; Hughson et al. 2022). Following impacts on Ceres, subsurface temperatures substantially exceed 273 K, so liquid water and brines can flow along pressure gradients generated by progressive freezing and surface topography (Bowling et al., 2019; Schmidt et al., 2020). The Occator Crater impact likely generated a large transient brine reservoir extending kilometers below the surface and capable of facilitating the ascent of over-pressurized brines from the mantle, as well as a tens to hundreds of meters thick brine-rich impact melt deposit draped on the nascent crater topography (Schmidt et al. 2020; Bowling et al. 2019; Hesse & Castillo-Rogez 2019; Quick et al. 2019). Although there is substantial interest in the possibility of deep mantle brine extrusion inside Occator Crater, Schmidt et al. (2017, 2020) hypothesized that formation of domes and hills on the crater floor was driven primarily by the shallow impact melt freezing process. In this proposed scenario, deep brine reservoirs at tens of kilometers depths and hydrologic connections between the crust and mantle are not necessary to explain the morphological characteristics of the Occator floor.

Here, we test the hypothesis that freezeout of impact melt deposits on the floor of Occator Crater may generate and transport enough water to create pingos on the surface, without the need of deep crustal brine ascent. Hornum et al. (2020) simulated terrestrial pingo formation conditions in Svalbard, Norway, by employing coupled 1D thermal and 3D groundwater flow models. They argued that a hybrid process of the classic hydraulic and hydrostatic scenarios can feasibly form terrestrial pingos. In their treatment, freezing front aggradation and degradation mirrored variations in mean annual air temperatures over thousands of years and expelled and transported pore water to the near surface to grow pingos. This is a hybrid form of the classic hydraulic and hydrostatic models, because the source of water is from aggrading permafrost (like the hydrostatic model), but individual pingo systems are not confined within frozen ground and hydraulic pressure from topography also drives groundwater flow (like the hydraulic model). This process is directly analogous to impact melt freezeout in Occator Crater. We have adapted Hornum et al. (2020)'s numerical modeling approach to assess the plausibility of pingo formation on Ceres.

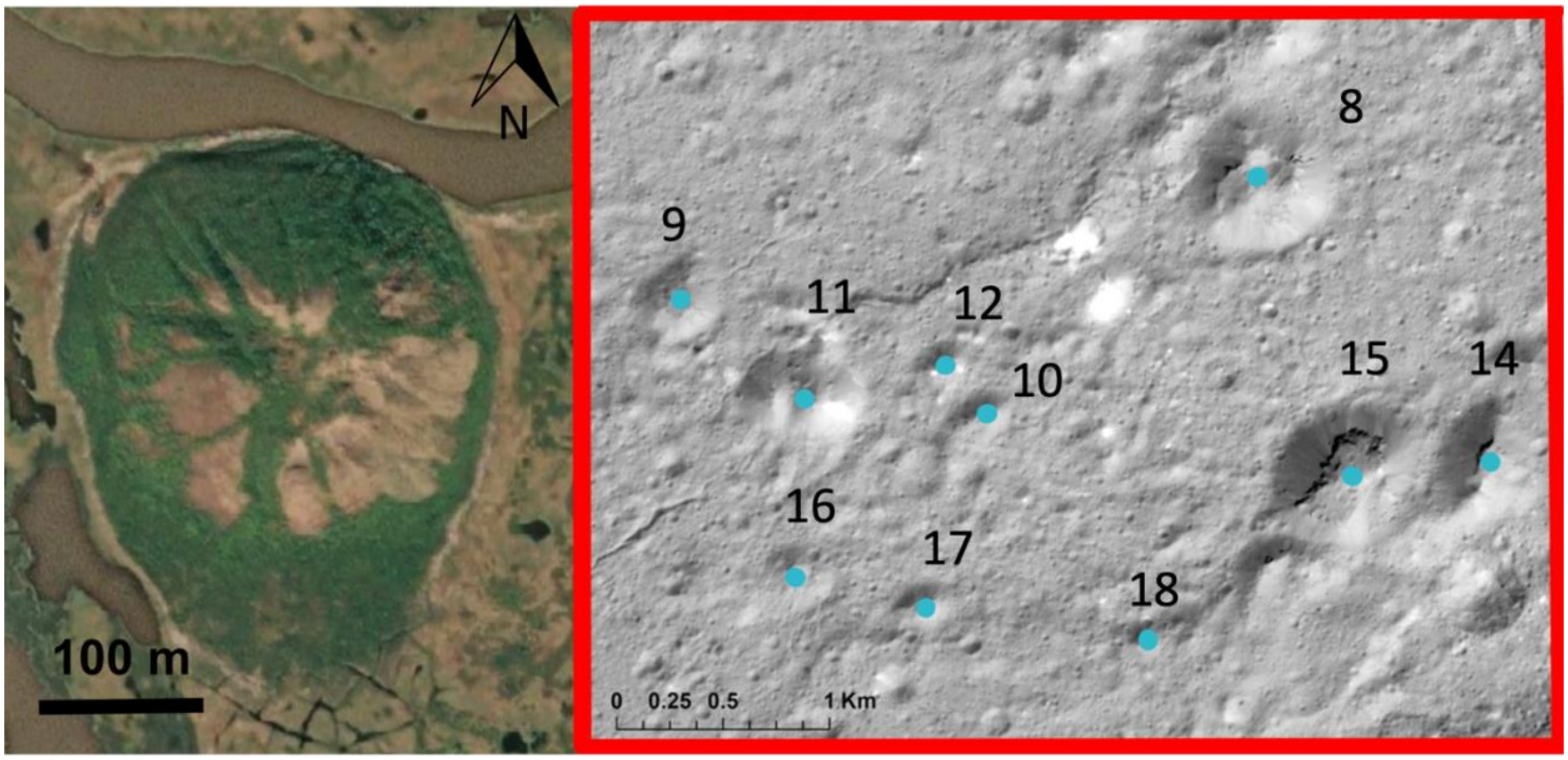

**Figure 1.** Left) Ibyuk pingo in Tuktoyaktuk, Northwest Territories, Canada; Right) Pingo candidates on the floor of Occator Crater, Ceres (blue) and their numeric identifiers. Imagery is from the XM2 clear filter FC mosaic (~3 m/pix) (Roatsch et al., 2018). Note similar planform shapes and ruptured summit morphologies. Figure 2 shows these pingo candidates in context with surrounding geology.

We simulated the evolution of a contiguous impact melt sheet in the southeastern portion of the floor of Occator Crater, where there is a cluster of 20 pingo candidates with strong morphologic similarity to terrestrial features (Fig. 2). We explored a comprehensive parameter space, characterizing the local impact melt sheet across the likely range of initial porosity, permeability, salt concentration, compressibility, and local geothermal gradient to investigate which combinations of these properties, if any, yield subsurface water transport and pingo growth on the surface.

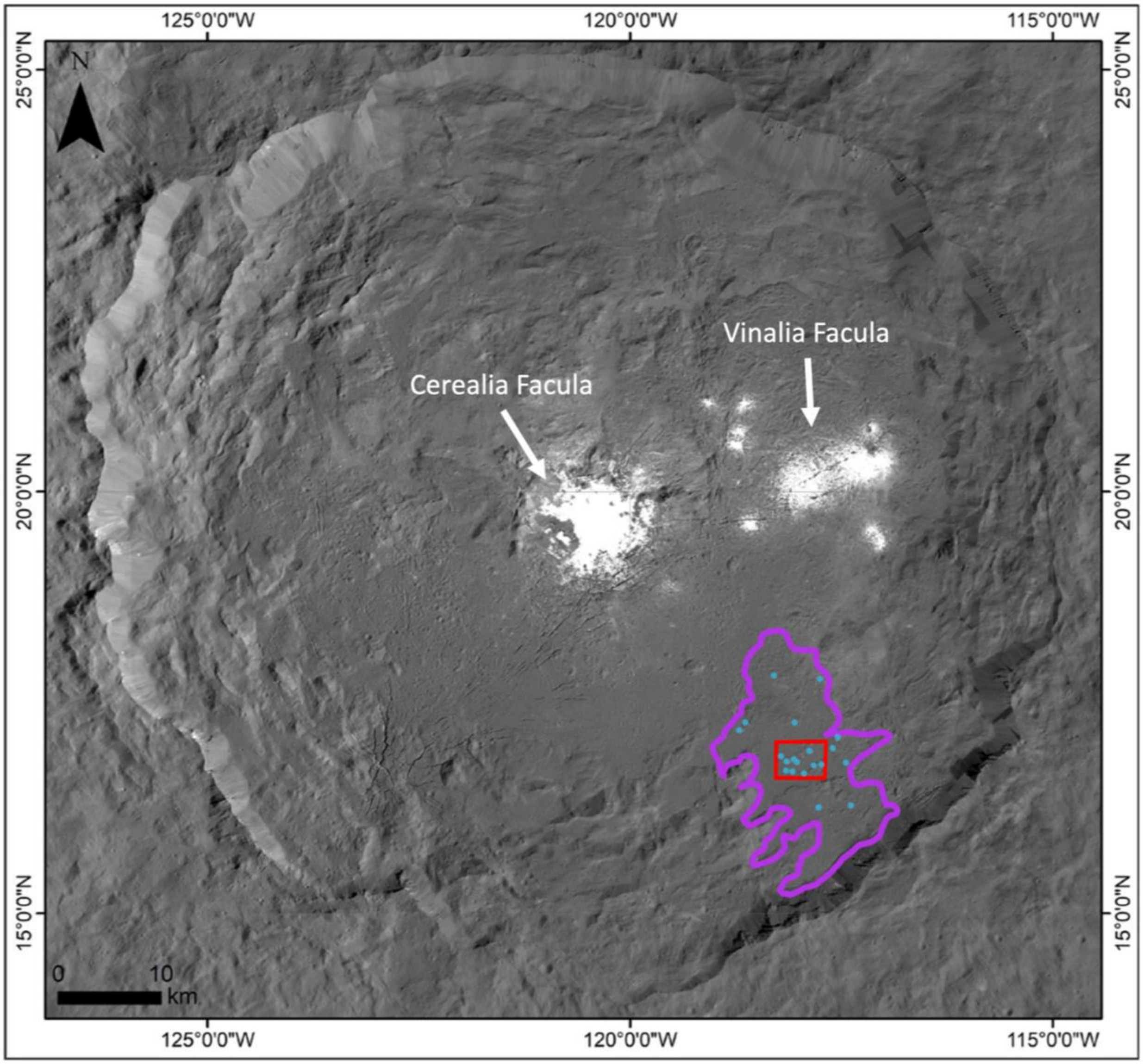

**Figure 2.** Occator Crater (19.82 ºN, 120.67 ºW) with our study region and groundwater flow model domain shown in purple and 20 pingo candidates marked in blue. Red box encompasses pingo candidates shown in figure 1. Occator Crater is 92 km across, and faculae on the floor are salt deposits left behind from surface extrusion of brines. The basemap is a combination of the XM2 clear filter FC mosaic (~3 m/pix) (Roatsch et al., 2018) the ~10 m/pixel XM2 clear filter FC controlled mosaic, and the LAMO clear filter FC controlled mosaic (~35 m/pixel) (Roatsch et al., 2017). Figure adapted from Scully et al. (2020).

We describe the geologic context of our simulations in section 2, thermal and groundwater flow model parameters and simplifying assumptions in section 3, results in section 4, discussion and implications of results in section 5, and conclusions in section 6.

**2 Geologic Context in Occator & Numerical Model Domain**

The key assumptions of our conceptual model of cerean pingo formation are that: (1) The relevant setting is sheets of impact melt on the floor of young craters. In Occator Crater, impact melt facies are typically 200–600 m thick (Scully et al. 2020); the stratigraphy of Ceres' older flat-floored craters is similar (Sizemore et al. 2017; 2019). (2) Melt sheets are deposited as mud slurries, consisting primarily of clay minerals, salts, water, and ice (Scully et al. 2019; Schenk et al. 2020). (3) Immediately after deposition, bulk flow of the slurry is common. Thin chill crusts form and are disrupted during the bulk flow stage, resulting in a characteristic ropey, pahoehoe-like texture in portions of Occator Crater (Scully et al. 2019; Schenk et al. 2020). (4) After sufficient cooling, bulk flow of the melt sheet terminates, but flow of brines through the underlying rock matrix can persist under a thickening ice-cemented chill crust (Ruesch et al. 2019; Quick et al. 2019). (5) The progressive, top-down freezeout of the remaining liquid can drive groundwater flow in direct analogy to the process by which permafrost aggradation drives pingo growth on Earth.

To capture the maximum extent of impact melt thicknesses reported by Scully et al. (2020), we defined a thermal model domain 600 m deep with 300 computational cells of 2 m height. Each one-dimensional model column was mapped to a 100 m x 100 m area (i.e., the lateral extent of a cell in the groundwater flow model) to cover our study region (Figs. 2 & 3). The physical dimensions of our model domain are similar to model domains for terrestrial pingo growth simulations, while the composition and temperature regime are quite different. Our thermal model domain is intermediate between two previous classes of theoretical studies of the thermal behavior of Ceres: (1) shallow ice retention studies relevant to the top centimeters to meters of the subsurface (e.g., Schorghofer, 2009; Schorghofer et al. 2016; Hayne & Aharonson 2015; Titus, 2015; Landis et al. 2017; etc.) and (2) deep interior studies relevant to the differentiation of Ceres and the thermal and rheological behavior of the crust following impacts, spanning depth scales of a few to hundreds of kilometers (e.g., McCord & Castillo-Rogez 2018; Castillo-Rogez et al. 2019; King et al. 2018; Bland et al. 2019; Hesse & Castillo-Rogez 2019; etc.). These two families of prior studies inform the initial and boundary conditions for our model, as well as our choices of material properties, which are a hybrid of the two endmembers.

Hesse & Castillo-Rogez (2019) modeled post-impact cooling in the vicinity of an Occator-sized impact (depth scale of tens of kilometers). In their domain, water ice, clathrates, and salts were, together, volumetrically dominant over clay minerals. Our simulation space—a freezing melt sheet 600 m thick—connects the deep region simulated by Hesse & Castillo-Rogez (2019) and the shallow subsurface considered in ice-retention simulations (cm to m; e.g., Schorghofer, 2016; Landis et al., 2017; 2019; 2025; Prettyman et al., 2021), where clay minerals are volumetrically dominant and water ice may not fully saturate the pore space. We assumed that groundwater flow relevant to pingo growth occurs only after bulk motion of the melt sheet has ceased, a mechanically stable, decimeter to meters-thick chill crust comprised of ice-cemented silicates and salts has formed (Ruesch et al. 2019; Quick et al. 2019), and a centimeters-thick dry lag controls both the average surface temperature and a vanishingly low rate of $H_2O$ sublimation to space (Schorghofer, 2016; Prettyman et al., 2021). The dry regolith lag is an efficient insulator that dampens the thermal wave and shields the impact melt beneath from seasonal thermal oscillations. We assumed the initial slurry was composed of particulate clays (porosity 25%–55%), saturated with liquid water/brine or locally cemented by water ice, with soluble salts making a minor volumetric contribution. These choices are intermediate between those made for Ceres at depth-scales of 1–20 km by Hesse & Castillo-Rogez (2019) and the typical assumption made in Ceres ice-retention studies (e.g., Scorghofer, 2016; Landis et al., 2017; 2019; 2025; Prettyman et al., 2021) that rock comprises 45–80% of the regolith by volume in the uppermost 1–5 m of the subsurface.

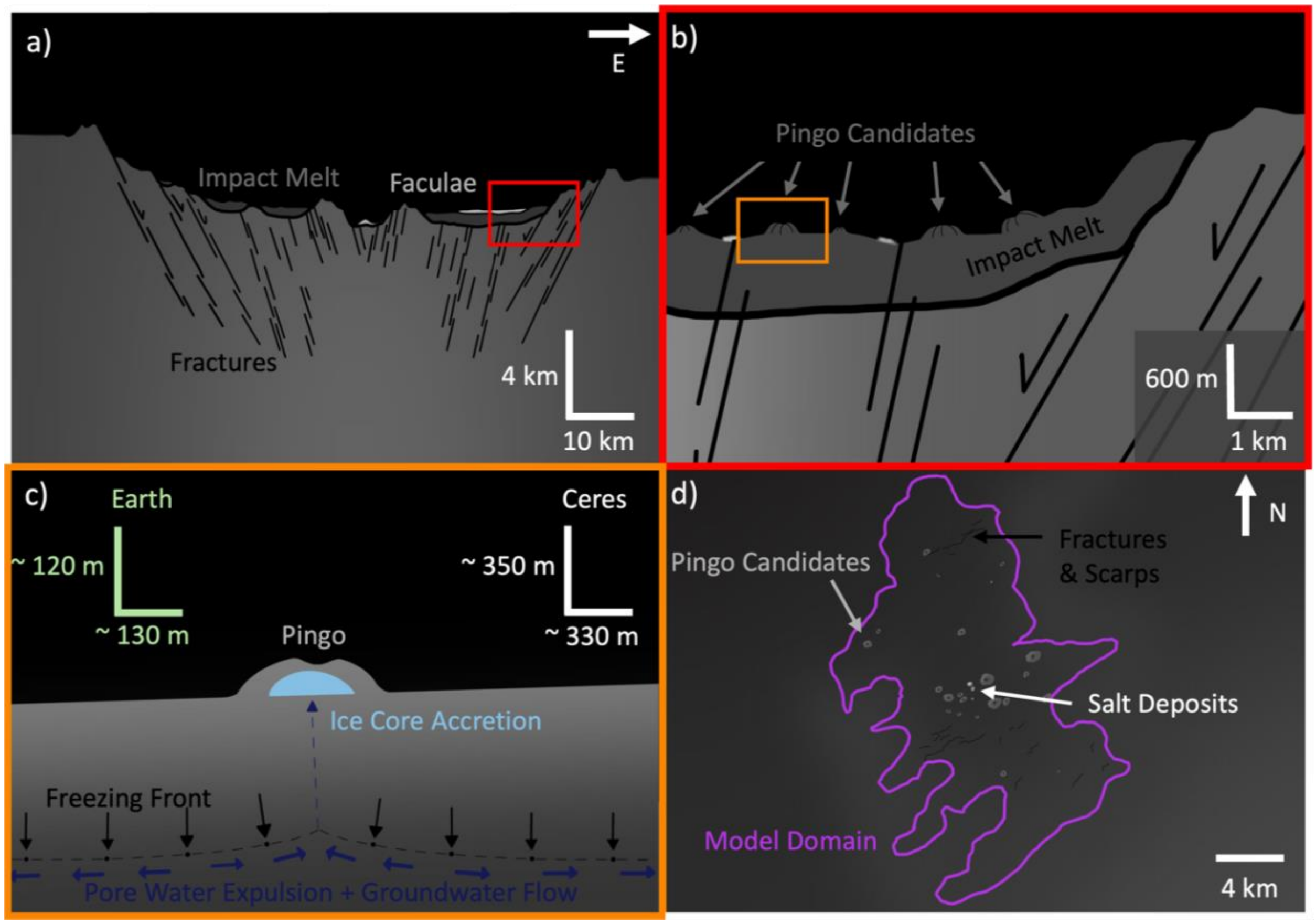

**Figure 3.** a) Cross-section of Occator Crater, adapted from Scully et al. (2020). Note ~600 m thick impact melt sheets draped over crater topography and faculae on top; b) Cross-section of a portion of impact melt with fractures, salt deposits, and pingo candidate hills on the surface; c) Cross-section of an idealized pingo atop impact melt, with additional scale bars for Earth as a comparison. Scale bars are based on the largest pingo and pingo-candidate on Earth and Ceres, respectively. In our model, freezing front aggradation leads to pore water expulsion and groundwater flow that may accrete onto near-surface pingo ice cores. Figure adapted from Hornum et al. (2020); d) Top-down view of our groundwater flow model domain: a contiguous impact melt sheet in the southeastern quadrant of Occator Crater. Key geomorphology includes fractures, salt deposits, and pingo candidates.

Our conceptual model of melt sheet freezing directly parallels the treatment of terrestrial pingos in Svalbard by Hornum et al. (2020). Our study region is a sheet of impact melt draped over underlying topography, and the lateral extent of the domain is much larger than its depth. Likewise, the Svalbard region considered by Hornum et al. (2020) is comprised of a relatively thin (hundreds of meters) pervious layer that conforms to local topography over a region of many square kilometers. The Svalbard model assumes an initial permafrost-free subsurface. An abrupt negative shift in surface energy balance initiates freezing front aggradation downward. Pressure develops at the freezing front, leading to pore water expulsion and hydraulic head gradients (Mackay 1979; Hornum et al. 2020). Fracturing or variable subsurface permeabilities may create localized discontinuities in the permafrost that allow pressurized groundwater to flow toward the surface (Hornum et al. 2020; Scheidegger et al. 2012). We—and Hornum et al. 2020—assume that such discontinuities exist at observed pingo locations and test whether recharge from permafrost aggradation can generate sufficient groundwater flow volumes through said discontinuities to form the observed pingos. In this scenario, groundwater flow has two main pressure sources: hydrostatic pressure from aggrading permafrost and hydraulic pressure generated by surface topography.

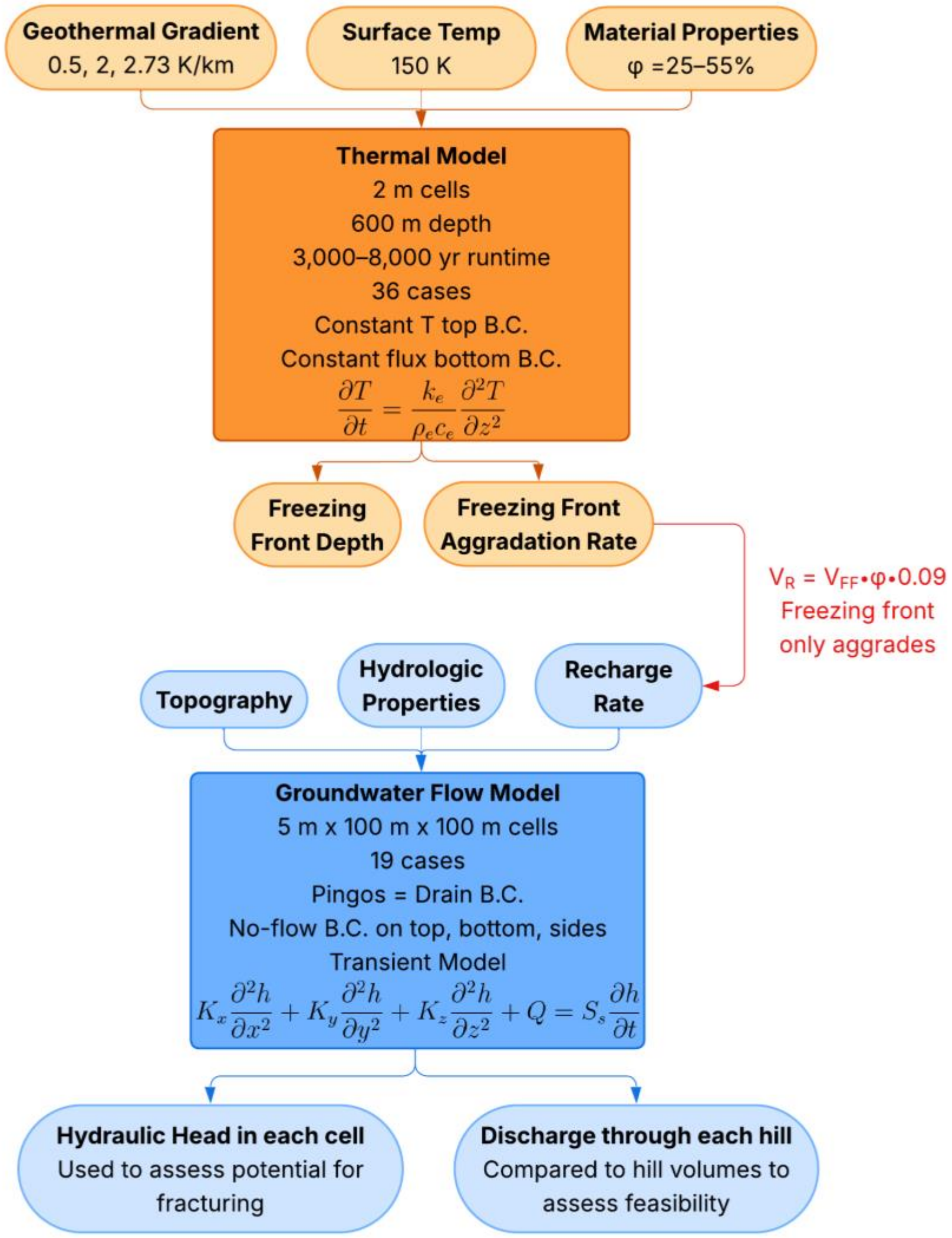

**Figure 4.** Flowchart showing inputs, model setup information, and outputs for thermal (orange) and groundwater flow (blue) models. Freezing front aggradation rate is mapped to recharge rate by the equation in red. See sections 3.1.3 for an in-depth discussion of our input parameters.

## 3 Numerical Models

### 3.1 Thermal Model

We directly adapted the one-dimensional conductive heat transfer model used by Hornum et al. (2020) to our study. The model solves an explicit forward-difference time approximation of the one-dimensional heat conduction equation:

$$\frac{\partial T}{\partial t} = \frac{k_e}{\rho_e c_e} \frac{\partial^2 T}{\partial z^2} \tag{1}$$

where $T$ is temperature [K], $t$ is time [s], $z$ is distance [m], and $k_e$, $\rho_e$, and $c_e$ are the effective values of the thermal conductivity [W (m K)$^{-1}$], density [kg m$^{-3}$] and specific heat capacity [J (kg K)$^{-1}$], respectively. We neglected advection because the low-temperature and low-hydraulic conductivity environment yield very small groundwater velocities such that conduction dominates over advection (see Supplement for discussion of Péclet Number; Bergman 2011). Hornum et al. (2020) simulated the thermal evolution of permafrost in Svalbard, Norway, during the time since the last glacial maximum, with historical climate records providing the surface temperature boundary condition over the past several thousand years. The bottom of their domain was set by a constant flux boundary condition, for which Hornum et al. (2020) used 0.025 ºC/m. We adapted this thermal model to Ceres by defining material properties and initial and boundary conditions consistent with our conceptual model of a cooling impact melt deposit inside Occator Crater.

#### 3.1.1 Material Properties

Lobate impact melt deposits inside Occator are thought to be broadly similar to Ceres' bulk crustal composition, which is rich in silicates, salts, clathrates, and ice (Schenk et al., 2020). Clathrates destabilize upon impact and are not expected to re-form upon freezing due to low pressures in the depth range of our model domain (< 2 km; Mousis et al. 2015; Sizemore et al. 2019), a distinct difference between the thermal characteristics of our study region and that of, for example, Hesse & Castillo-Rogez (2019). We calculated effective values for bulk material properties of the impact slurry and chill crust based on the volumetric fraction of the constituent material phases (i.e., mineral fragments, solid ice, and liquid water). We defined the total lithic fraction to depend solely on the input porosity, $\phi$. The partitioning of the volatile components (water vs. ice) in the pore space was described by "soil freezing curves" (SFCs), which characterize how the pore liquid fraction varies with temperature as the saturated regolith freezes (see section 3.1.2 below). The effective thermal conductivity was calculated as the root mean square as done by Mottaghy and Rath (2006), Govaerts et al. (2016), and Hornum et al. (2020):

$$k_{eq} = \left(f_s\sqrt{k_s} + f_i\sqrt{k_i} + f_w\sqrt{k_w}\right)^2 \quad (2)$$

where $f$ denotes the volume fraction of a phase in the pore space, and the *s, w,* and *i* subscripts denote soil/rock, water, and ice, respectively. Latent heat was incorporated into the model through calculation of the effective volumetric heat capacity:

$$C_{eq} = c_e\rho_e = F_s\rho_s c_s + F_w\rho_w\left(c_w + \frac{\partial f_w}{\partial T}L\right) + F_i\rho_i\left(c_i + \frac{\partial f_w}{\partial T}L\right) \quad (3)$$

where $F$ denotes the total fraction of a phase and $L$ is the latent heat of fusion [J kg$^{-1}$] (Hornum et al. 2020; Govaerts et al. 2016; Mottaghy & Rath 2006). Table 1 lists material properties used for our model inputs. Thermal dynamics are dependent on the material properties of the medium, and these values are unconstrained for cerean impact melts. Our chosen properties are consistent with Hesse & Castillo-Rogez (2019), who modeled the thermal evolution of the post-impact environment at Occator Crater. The influence of material properties on impact melt cooling dynamics on Ceres merits further study.

**Table 1.** Material properties in thermal model.

| Material | k [W (m K)$^{-1}$] | ρ [kg m$^{-3}$] | $c_p$ [J (kg K)$^{-1}$] | L [J kg$^{-1}$] | Source(s) |
|---|---|---|---|---|---|
| Rock | 2 | 2430 | 2000 | - | Hesse & Castillo-Rogez (2019); Grindrod et al. (2008) |
| Liquid Water | 0.56 | 1000 | 4180 | - | Williams & Smith (1989) |
| Ice | 2.24 | 917 | 2100 | 3.336 x 10$^5$ | Williams & Smith (1989); Mottaghy & Rath (2006) |

3.1.2 Soil Freezing Curves

An important aspect of permafrost dynamics is the persistence of substantial liquid $H_2O$ in the pore space of a soil, regolith, or porous bedrock below the bulk freezing temperature of 273 K (e.g., Ostroumov and Siegert, 1996; Dash et al., 2006). This "residual liquid" is produced by both pore-space confinement effects and by solute-induced freezing point depression (e.g., Dash et al., 2006; Chen et al. 2020). Substantial progress has been made in the characterization of residual liquid as a function of temperature for specific combinations of granular materials and solutes over the past two decades (e.g., Jakosky et al., 2003; Rempel, 2008; Sizemore et al., 2015; Chen et al., 2020; Lei et al., 2022; Sizemore et al., 2026). The bulk thermal properties of our model domain depend strongly on the partitioning of the liquid and soil phases in the pore space, so selecting a range of SFCs relevant to Ceres was a critical aspect of defining the larger parameter space for our investigation.

Many of the characteristics of the Occator impact melt relevant to the definition of an SFC are poorly constrained, including porosity, grain size distribution, and the relative and bulk concentration of various salt species. Variations in regolith grain and pore geometry—as well as the presence of multiple salt species with distinct eutectic temperatures—may produce complex SFCs at the low temperatures relevant to Ceres (Chen et al., 2020; Sizemore et al., 2026). However, the broad character of soil freezing curves has been well-described for almost three decades (e.g., Andersland and Ladanyi, 2001). In this context, we have chosen two endmember SFCs to span the full range of likely freezing behaviors in Ceres' shallow subsurface, as well as a family of intermediate best-approximation SFCs to complete our parameter space. We assumed that the Occator impact melt sheet initially contained ~0–15% soluble salts by volume. We then adopted representative salt-free (rapid-freezing), high-salt (slow-freezing), and intermediate (clay-dominated) SFCs from the literature (Figure 5).

We adopted our "salt-free" SFC from Hornum et al. (2020). They made an ad hoc assumption of a smooth SFC with rapid freezeout between 273.15 K and a solidus temperature of 271.15 K to represent the freezing behavior of the talik beneath the Svalbard pingos. Their stated rationale for choosing a smoothed step function was to improve convergence of their numerical model, and they argued that the shape of the freezing curve was arbitrary. However, their specific choice was also consistent with empirical measurements and theoretical predictions of SFCs in salt-free Arctic permafrost materials. Specifically, the Hornum et al. (2020) endmember is consistent with families of SFCs developed for salt-free soils dominated by silt- to sand-sized grains (Andersland and Ladanyi, 2001; Chen et al., 2020; Sizemore et al., 2015; 2026).

We adopted our "high-salt" SFC from Hesse & Castillo-Rogez (2019), who also developed an ad hoc curve appropriate for a system with large quantities of salt. They assumed a linear ramp between bulk melting at the liquidus temperature (273.15 K) and total freeze-out at an effective eutectic temperature for a system with multiple salt species, which Castillo-Rogez et al. (2018) previously estimated at 245 K based on the likely composition of Ceres' crust. Hesse & Castillo-Rogez (2019) assumed Ceres' crust was composed of 15% salt, 20% silicate rock, and variable amounts of ice and clathrates. Their SFC represents an extreme slow-freezing scenario for our shallower model domain (46 km vs. 600 m), which we assumed to be volumetrically richer in silicate rock and poorer in $H_2O$. This is due to the range of clay porosities tested (see section 3.1.3) and the inability of clathrates to re-form post-impact (Mousis et al. 2015; Sizemore et al. 2019).

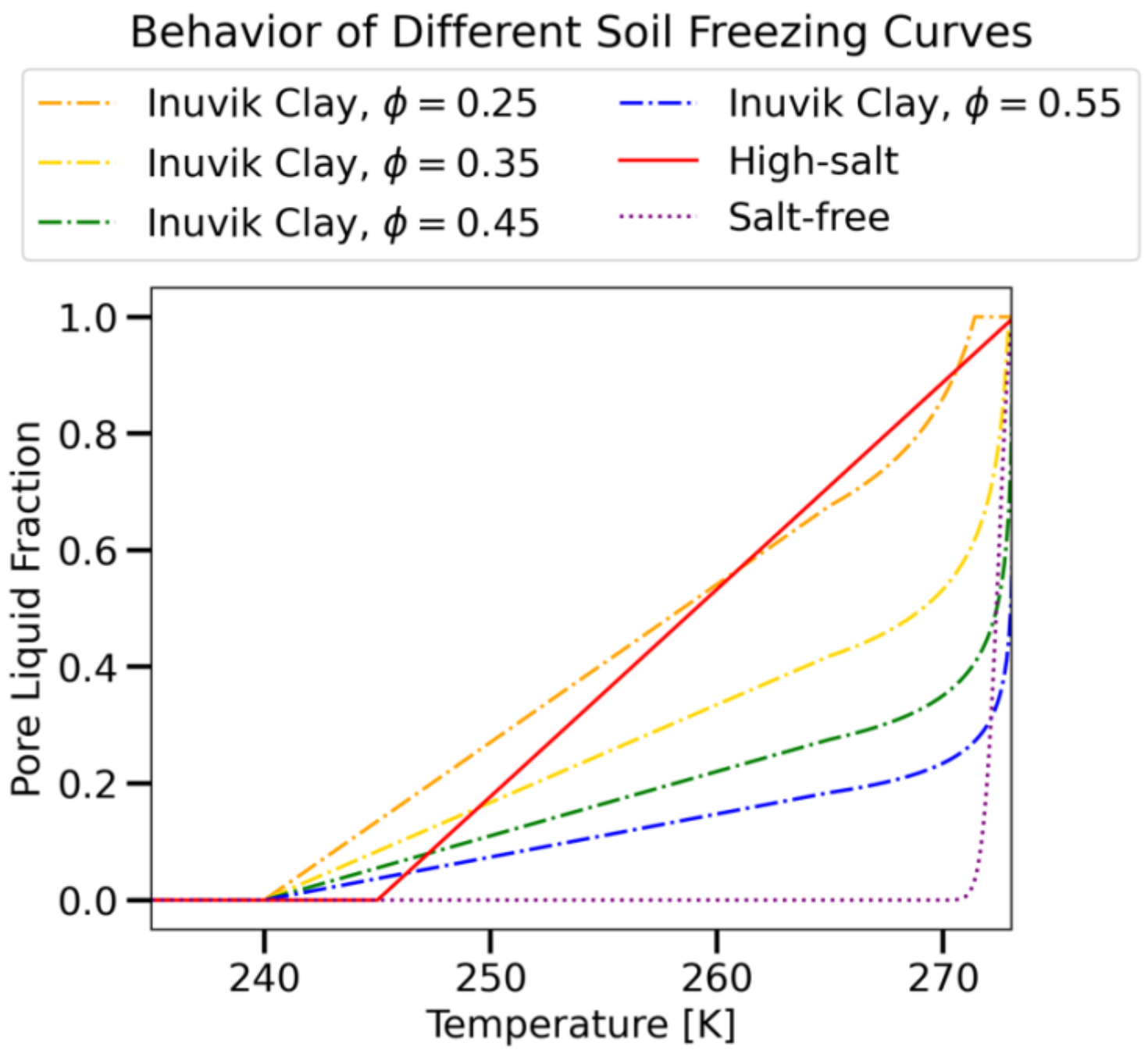

**Figure 5.** Pore liquid fraction vs. temperature for the three soil freezing curves (SFCs) used in our model. Salt-free assumes zero salt in the pore space. High-salt assumes the volume of the phase space is 15% salt. The family of Inuvik Clay SFCs are porosity-dependent and developed from laboratory data (Andersland & Ladanyi 2001; Rempel 2007; 2008; Sizemore et al. 2015).

We selected a family of intermediate SFCs from the soil-freezing literature to represent our best approximation of the freezing behavior of clay-dominated impact slurries on Ceres based on their composition and observed properties. We refer to this SFC family as "Inuvik Clay," based on soil from Inuvik, Northwest Territories, Canada, a region of widespread permafrost from which the curves were derived. We refer to it as a "family" of SFCs, with each porosity having a different curve despite them all being from Inuvik Clay. The original power law parameterization of the Inuvik Clay SFC was reported by Andersland and Ladanyi (2001) in a larger catalog of terrestrial SFCs, and later adapted by Rempel (2007, 2008) and Sizemore et al. (2015) for application to terrestrial and martian permafrost models, respectively. Since we assumed impact melt deposits are slurries of clay minerals, salts, water, and water ice, Inuvik Clay is an appropriate choice to represent the particle geometries and storage capacity we might expect in the Occator melt. Additionally, for regolith dominated by clay-sized particles at temperatures significantly below bulk freezing, pore-space confinement effects dominate over colligative effects (i.e., freezing point depression due to salts), making clay SFCs relatively insensitive to solute content (Sizemore et al. 2015; Chen et al., 2020; Lei et al. 2022). Thus, we specifically adopted the family of Inuvik Clay SFCs reported by Sizemore et al. (2015), which includes a dependence on assumed porosity that we exploit in our parameter studies. The simple log-normal parameterization reported by Sizemore et al. (2015) was based on high temperature (> -10° C) terrestrial data and does not lead to total freezeout even at the extreme low temperatures of Ceres (i.e., pore liquid fraction asymptotically approaches 20% at 250 K). We thus forced total freezeout by modifying the SFCs to include a linear ramp between 265 K and 240 K. We chose these temperatures to achieve a smooth transition between the slope of the log-normal SFCs and an artificially imposed solidus temperature, following the approach of Hesse & Castillo-Rogez (2019). This approach is also broadly consistent with qualitative discussion of the shape of multi-solute SFCs by Sizemore et al. (2026), who studied the geometric characteristics of thin films and veins in martian permafrost. A liquid with multiple solutes has multiple eutectic temperatures, corresponding to steps in the SFC. Hesse & Castillo-Rogez (2019) assume that a multi-solute brine has many eutectic steps that can be approximated by a smooth curve.

#### 3.1.3 Thermal Model Boundary Conditions & Parameter Space

We imposed an initial temperature profile of 273.15 K throughout our model domain, representing the point at which bulk flow of the impact melt has ceased and fluid is allowed to freeze and flow along pressure gradients. We set the temperature of the top model cell to 150 K (i.e., the mean annual surface temperature of Ceres at Occator Crater's latitude; Hesse & Castillo-Rogez, 2019; Bowling et al., 2019), and the temperature of the bottom model cell was dictated by a constant flux condition set by the geothermal gradient. The temperatures of the model cells then evolved according to Eq. 1. Our choice of 2 m cell thicknesses allowed us to neglect diurnal and seasonal temperature oscillations because temperatures beneath the diurnal skin depth of the surficial dry regolith lag layer (see Table 2) remain near the mean annual surface temperature (Rivkin et al. 2011; Hayne & Aharonson 2015; Landis et al. 2017). Additionally, the scale of seasonal oscillations (meters) is trivial compared to the scale of our model domain (hundreds of meters). It is also consistent with choices made by Hornum et al. (2020) to optimize numerical efficiency. We ran each model for 10,000 Earth years with a timestep of 0.05 years. Hornum et al. (2020) also used this time step, and it is within the stability criteria for model convergence.

We tested three SFCs, four porosities (25%, 35%, 45% and 55%), and three geothermal gradients (0.5 K/km, Bowling et al. 2019; 2 K/km, Hesse & Castillo-Rogez 2019; and 2.73 K/km, Raymond et al. 2020) for a total of 36 thermal model runs. We chose these porosities because (1) they span the range of measured porosities of clays and silts (Fetter 2018; Sizemore et al. 2015); (2) they are in the range of ice abundances used in Hesse & Castillo-Rogez (2019); and (3) terrestrial permafrost lag deposits may have average porosities up to ~68% or higher, and we wanted to capture the extremes of what is observed in analogous terrestrial environments (Sizemore & Mellon 2008; Douglas & Mellon 2019). Bland et al. (2016) suggested that the cerean subsurface is likely 30–40% ice by volume with 60–70% comprised of rocks, salts, and clathrates, but higher porosities are plausible in impact melt deposits considering the lack of clathrates and that the pore space is occupied by liquid water, salt, and ice—as opposed to ice alone. Additionally, work by Palmer et al (2021) suggested that cerean regolith may be more porous than previously expected, with a bulk porosity ranging from ~50–70%. Pamerleau et al. (2024) found that Ceres' crust may be up to 90% ice and still agree with Dawn observations. Since the composition of the impact melt is representative of the crust (Schenk et al. 2020), bulk properties of the crust are informative for our model. Sizemore et al. (2019) also showed that ice concentration (i.e., porosity, in our treatment) is heterogeneous both vertically and laterally on length scales from < 1 km to hundreds of km. Outputs of our model included temperatures of all model cells at the end of each year (i.e., 20 iterations). From these temperatures, we calculated the freezing front depth within the model domain using the solidus temperature of each SFC. Running our model for 10,000 years allowed us to calculate freezing front aggradation rates (i.e., how many meters the freezing front advanced each year) which we used to solve for the equivalent recharge rates that set the initial conditions for the groundwater flow model (Figure 4).

**Table 2.** Thermal diffusivity, thermal inertia, and diurnal and annual skin depths for impact melt compositional endmembers and the surficial dry regolith layer.

| Material | Thermal Diffusivity [$m^2$/yr] | Thermal Inertia [tiu] | Diurnal Skin Depth [m] | Annual Skin Depth [m] |
|---|---|---|---|---|
| Φ = 25%; 0% liquid; 100% ice | 15.73 | 2914.65 | 0.07 | 4.8 |
| Φ = 25%; 100% liquid; 0% ice | 10.67 | 2677.06 | 0.06 | 3.95 |
| Φ = 55%; 0% liquid; 100% ice | 20.70 | 2629.7 | 0.08 | 5.5 |
| Φ = 55%; 100% liquid; 0% ice | 15.64 | 3024.91 | 0.07 | 4.78 |
| Dry Regolith Layer (Landis et al. 2017; Rivkin et al. 2011) | - | 15 | ~0.001–0.003 | ~0.067–0.2 |

### 3.2 Groundwater Flow Model

Following Hornum et al. (2020), we used the USGS MODFLOW code to simulate groundwater flow within freezing impact melt deposits on the floor of Occator Crater (McDonald & Harbaugh 1988). MODFLOW is a three-dimensional finite difference code with both steady state and transient modeling capabilities. We used a transient model to integrate the storage capabilities of a clay aquifer.

#### 3.2.1 Groundwater Flow Model Domain

To determine which hills to target as pingo candidates in our model, we filtered mapped hills in impact melt deposits from Hughson et al. (2022). To be considered a pingo candidate, each mound must be (1) a distinct topographic feature; (2) round to oblong in planform; and (3) have distinct slope change at its base relative to the surrounding terrain. We chose to model an impact melt deposit in the southeastern quadrant of the crater floor near the crater rim because it has the highest concentration of morphologically compelling pingo candidates (Figure 2).

Model cells were 100 m x 100 m x 5 m. There were 120 cells in a column for a depth of 600 m. The surface area of the model domain was 210.78 km$^2$ (shown in Figure 2). We set the top elevations of our model domain with a primary mission-derived digital terrain model (DTM) of Occator Crater (NASA Planetary Data System, https://sbnarchive.psi.edu/pds3/dawn/fc/DWNCLSPG_2/, accessed 2023). Elevations span 1150 m of relief. See Supplement for a 3D rendering of our MODFLOW model domain. Since bedrock topography is unknown, we assumed that bedrock topography mirrors the surface topography. The thermal and groundwater flow models were coupled through the freezing front aggradation rate outputs of the thermal model. Assuming expelled pore water as the sole source of water in our groundwater flow model, we converted freezing front aggradation rates $V_{FF}$ [m/s] to equivalent recharge rates $V_R$ [m/s] via the porosity $\phi$ and volumetric expansion of water as it freezes $X_w$ (Figure 4):

$$V_R = V_{FF} * \phi * X_w \tag{4}$$

We converted the equivalent recharge rate to an areal recharge rate $Q_R$ [m$^3$/s] by multiplying $V_R$ by the cross-sectional area of the cell $A$ [m$^2$]:

$$Q_R = V_R * A \tag{5}$$

We approximated pingo candidates in the model domain as drains with the MODFLOW Drain package. Drains removed water from the domain once the hydraulic head in the drain cell was greater than the assigned drain elevation (i.e., pingo elevation from the DEM).

#### 3.2.2 Material Properties

Groundwater flows through a three-dimensional porous medium according to Darcy's Law:

$$K_x \frac{\partial^2 h}{\partial x^2} + K_y \frac{\partial^2 h}{\partial y^2} + K_z \frac{\partial^2 h}{\partial z^2} + Q = S_s \frac{\partial h}{\partial t} \quad (6)$$

Where $K$ is hydraulic conductivity [m/s], $h$ is hydraulic head [m], $x$, $y$, and $z$ are distances [m], $Q$ is a source/sink term in units of volumetric flux per unit volume [$s^{-1}$], $S_s$ is specific storage [$m^{-1}$], and $t$ is time [s]. Water flows from high to low hydraulic head, and the rate of discharge is modulated by the hydraulic conductivity. In our model, $Q$ can be positive (recharge from freezing front aggradation) or negative (pingo drains) depending on the cell. Hydraulic conductivity depends on both material and fluid properties:

$$K = \frac{\rho_f g \mathrm{k}}{\mu} \quad (7)$$

Where $\rho_f$ is density of the fluid [kg $m^{-3}$], $g$ is acceleration due to gravity [m $s^{-2}$], $k$ is permeability of the material [$m^2$], and $\mu$ is fluid viscosity [Pa s]. We used commonly used densities and viscosities of water (1000 kg $m^{-3}$ and $10^{-3}$ Pa s, respectively) and an acceleration due to gravity of 0.27 m $s^{-2}$. Permeabilities of planetary materials are poorly constrained and can span many orders of magnitude. Permeability depends on factors like porosity, particle size, particle geometry, and the presence of impurities. We used permeability values consistent with clays of different porosities (Table 3). To explore a wider range of permeabilities, we ran 4 additional groundwater flow model runs using average lab-measured permeabilities of glass spheres, Mars-analog regolith, dune sand, loess, and Antarctic Dry Valley soils from Sizemore & Mellon (2008). Ceres' gravity is a fraction of Earth's, and its effects on compaction and permeability are poorly constrained. We extended our permeability space to account for these potential effects.

**Table 3.** Porosity-permeability pairs used in groundwater flow models. The first four rows are representative of clays; the bottom four rows are our expanded permeability space.

| Porosity | Permeability [$m^2$] | Source |
|---|---|---|
| 0.25 | 1E-20 | Neuzil (2019) |
| 0.35 | 1.4E-19 | Rempel (2007) |
| 0.45 | 1.354E-16 | Sizemore et al. (2015) |
| 0.55 | 1.292E-15 | Sizemore et al. (2015) |
| 0.25 | 5.54E-14 | Sizemore & Mellon (2008) |
| 0.35 | 4.64E-13 | Sizemore & Mellon (2008) |
| 0.45 | 3.95E-12 | Sizemore & Mellon (2008) |
| 0.55 | 1.71E-13 | Sizemore & Mellon (2008) |

These permeabilities produced hydraulic conductivities ranging from 2.7E-15 to 1.07E-6 m/s. We calculated specific storage as:

$$S_s = \rho_f(\alpha + \phi\beta) \quad (8)$$

Where $\alpha$ is aquifer compressibility [$Pa^{-1}$] and $\beta$ is water compressibility [$Pa^{-1}$] (Singhal & Gupta 2010). We used compressibility values relevant to clay aquifers: $\alpha$ = $10^{-8}$ to $10^{-6}$ $Pa^{-1}$ (Freeze & Cherry 1979). We assumed $\beta$ = 4.7E-10 $Pa^{-1}$ (Singhal & Gupta 2010).

### 3.2.3 Groundwater Flow Boundary Conditions & Parameter Space

Out of the 36 total thermal model runs, we down-selected 15 cases to model groundwater flow. On one end of the parameter space, we modeled the minimum porosity (25%), minimum geothermal gradient (0.5 K/km), maximum specific storage (2.7E-4 $m^{-1}$), salt-free SFC case which yielded fastest total freezeout and minimum total water (Case 1). On the other end of the parameter space, we modeled the maximum porosity (55%), maximum geothermal gradient (2.73 K/km), minimum specific storage (2.77E-6 $m^{-1}$), high-salt SFC case which yielded slowest total freezeout and maximum total water (Case 15). We added four high-permeability cases (cases 16–19) to probe a larger permeability parameter space. Table 4 summarizes all groundwater flow model runs.

**Table 4.** All groundwater flow model runs with input parameters, sorted by increasing permeability and decreasing specific storage.

| Case | Porosity | Permeability [$m^2$] | Geothermal Gradient [K/km] | SFC | Specific Storage [$m^{-1}$] |
|---|---|---|---|---|---|
| 1 | 0.25 | 1.00E-20 | 0.5 | Salt-free | 2.70E-04 |
| 2 | 0.25 | 1.00E-20 | 0.5 | High-salt | 2.70E-04 |
| 3 | 0.25 | 1.00E-20 | 0.5 | Salt-free | 2.73E-06 |
| 4 | 0.35 | 1.40E-19 | 2 | Inuvik Clay | 2.70E-04 |
| 5 | 0.35 | 1.40E-19 | 2 | High-salt | 2.70E-05 |
| 6 | 0.35 | 1.40E-19 | 2 | High-salt | 2.74E-06 |
| 7 | 0.45 | 1.35E-16 | 2.73 | Inuvik Clay | 2.71E-05 |
| 8 | 0.45 | 1.35E-16 | 2 | High-salt | 2.71E-05 |
| 9 | 0.45 | 1.35E-16 | 2 | High-salt | 2.76E-06 |
| 10 | 0.45 | 1.35E-16 | 2 | Inuvik Clay | 2.76E-06 |
| 11 | 0.45 | 1.35E-16 | 2 | Salt-free | 2.76E-06 |
| 12 | 0.55 | 1.29E-15 | 2.73 | High-salt | 2.70E-04 |
| 13 | 0.55 | 1.29E-15 | 2.73 | Inuvik Clay | 2.71E-05 |
| 14 | 0.55 | 1.29E-15 | 2.73 | Salt-free | 2.77E-06 |
| 15 | 0.55 | 1.29E-15 | 2.73 | High-salt | 2.77E-06 |
| 16 | 0.25 | 5.54E-14 | 0.5 | Salt-free | 2.73E-06 |
| 17 | 0.55 | 1.71E-13 | 2.73 | High-salt | 2.77E-06 |
| 18 | 0.35 | 4.64E-13 | 2 | High-salt | 2.70E-05 |
| 19 | 0.45 | 3.95E-12 | 2.73 | Inuvik Clay | 2.71E-05 |

We initialized the groundwater flow model by setting the head in each cell to the ground surface elevation of the DEM. As the freezing front aggraded downward, we deactivated all cells above the freezing front depth, assuming the pore space was completely frozen and impervious to flow. Recharge was deposited in the layer directly beneath the freezing front. Pingo drains were also located in this layer, but their prescribed elevations were held constant at the top of the domain. We calculated drain conductance (i.e., the rate at which water discharges through drains) according to:

$$C = \frac{KA}{dl} \tag{9}$$

Where $C$ is conductance [$m^2/s$], $K$ is hydraulic conductivity [m/s], $A$ is cell area [$m^2$], and $dl$ is distance between cell nodes.

MODFLOW automatically assigns no-flow conditions to the boundaries of the model domain on all sides. MODFLOW 6 is not equipped to simulate an advancing freezing front, so we temporally discretized each model run into ~15–20 "time blocks" (depending on total freezeout time) where the freezing front aggradation rate was approximately constant. During each time block, the depth of the freezing front was constant. Each time block was comprised of smaller, computational timesteps according to the material properties of the domain. For each time block, we calculated the average recharge rate from thermal model outputs (Eq. 4) and applied it to the layer beneath the freezing front depth. We increased the freezing front depth at the end of each time block to initialize the next time block. To maintain coupling between subsequent time blocks, we set the initial head in each cell as the final head from the previous time block. Model outputs included hydraulic head in each cell, cell-by-cell flow volumes, and flow volumes from boundary conditions (e.g., recharge, drains). We summed water discharge through each pingo candidate "drain" after each time block.

We also used hydraulic head outputs to assess the potential for hydraulic fracturing. When pore pressure exceeds the overburden stress and tensile strength of the overlying materials, hydraulic fracturing may occur. Fractures can provide additional groundwater flow pathways and opportunities to relieve pressure. Tensile strengths of ice-cemented clays and permafrost are an active area of research (e.g., Xu et al. 2020; You et al. 2021), and the tensile strength of cerean impact melt is unknown. We assumed a tensile strength of 2 MPa consistent with measured values of frozen clays (You et al. 2021) and estimated values of martian permafrost (Mellon et al. 1997). For each cell in our domain, we calculated the overburden stress then used the pore pressure and assumed tensile strength to determine if the conditions for hydraulic fracture were met.

### 3.3 Implications of Simplifying Assumptions

Our model is the first of its kind to investigate pingo formation on an extraterrestrial body, and it includes some necessary simplifications and assumptions. We assumed a subsurface homogeneous in composition, porosity, and permeability due to a paucity of subsurface structural data in Occator Crater. We also did not incorporate a temperature-dependent permeability, because introducing it would have added computational complexity and uncertainty to our flow model. Future laboratory work investigating planetary-relevant permeabilities at low temperatures is essential to refine studies of hydrologic processes on Ceres and other icy bodies.

We have assumed constant, temperature- and solute-independent liquid viscosities in our groundwater flow simulations and our calculation of hydraulic conductivities. Data constraining the types and abundance of salt species in Ceres' shallow subsurface are limited, and brine viscosities measured at Ceres-relevant temperatures are unconstrained in the literature. Our assumption of a constant pure $H_2O$ viscosity likely produces higher flow rates than cold brines in the cerean environment, but quantifying the magnitude of this effect is beyond the scope of the current work. We also assumed temperature-independent heat capacities and thermal conductivities. The temperature dependencies of these material properties are not well-qualified in the literature, as there is a lack of empirical data on the behavior of brine-ice-clay mixtures at low temperatures.

Limited knowledge of the cerean subsurface also prevents us from incorporating variable lithologies or entrained boulders in our simulations, both of which are likely in an impact melt flow process. We also assume that the impact melt is a constant thickness (600 m) and that surface topography mirrors bedrock topography. Portions of the impact melt may be as a thin as 100–300 m, resulting in a factor of 2–6 excess water volume if the entire melt sheet is thinner than assumed. Geomorphological evidence supports our choice of 600 m and suggests that our study region in the southeastern quadrant of Occator is the thickest region of melt (Scully et al. 2020).

We applied the current mean annual surface temperature at Occator Crater as our upper boundary condition (150 K assuming Ceres' current axial tilt of 4º; Bowling et al. 2019). Ceres' obliquity oscillates on a period of 24.5 kyr with a maximum tilt of 19.6º (Ermakov et al. 2017). Obliquity variation likely produced changes in the mean annual surface temperature on timescales relevant to impact melt freezing at Occator, in a manner analogous to $10^3$–$10^4$-year surface temperature variations modeled by Hornum et al. (2020) for Svalbard. We have not accounted for obliquity-driven surface temperature changes in our model, because there is no obliquity at which we expect the upper boundary condition to exceed or closely approach the bulk melting temperature of 273 K. Defining a surface temperature boundary condition based on dynamical models of Ceres' obliquity evolution is an area for possible future model refinement.

We assumed that all water that flowed into a pingo candidate "drain" cell accreted onto a growing ice core if the head in that cell was greater than the elevation of the hill. This approach likely overestimates water volume and neglects additional possible pathways for groundwater flow, like subsurface fractures (see section 5.1). Our handling of freezing front propagation in MODFLOW "turned off" several cell layers at a time when advancing the time block, effectively removing the water in those layers and underestimating the volume of water in the system. We assume this pair of over- and under-estimations approximately cancel.

## 4 Results

### 4.1 Thermal Simulations

Figure 6 shows the a) thermal and b) pore liquid fraction evolution of the model domain during the freezing process for three material regimes along with the c) freezing front aggradation rates and corresponding recharge rates. Total freezing times ranged from ~3000–8000 yr, depending on the geothermal gradient, porosity, and SFC used. The salt-free SFC, 25% porosity, and 0.5 K/km geothermal gradient case on the left of Figure 6a represents the fastest freezing case of our thermal simulations. The freezing front propagated quickly in this case, compared to other SFCs with lower eutectics. The high-salt SFC, 55% porosity, 2.73 K/km geothermal gradient case yielded the longest freezing time. With a eutectic of 245 K, water remained liquid in the pore space at lower temperatures due to colligative effects. The Inuvik Clay SFC, 45% porosity, 2 K/km geothermal gradient case demonstrated intermediate behavior, despite having the lowest "eutectic" temperature.

The evolution of the pore space is shown in Figure 6b. Ice growth in the pore space causes pore water expulsion, so the temporal evolution of the pore liquid fraction had important implications for the timing and amount of recharge. The salt-free SFC case underwent a stark transition from fully liquid-saturated to fully ice-saturated and the high-salt SFC case underwent a gradual transition. Inuvik Clay, intermediate between the two, underwent rapid freezing to ~265 K (about 90% liquid) and then froze more gradually. This was largely modulated by the shape of the SFC. Concave SFCs, like salt-free and Inuvik Clay from 273 K to 265 K (Figure 5), displayed rapid freezing behavior. The high-salt SFC, which is a linear ramp down to the eutectic, maintained a constant rate of ice growth for every degree of temperature decrease.

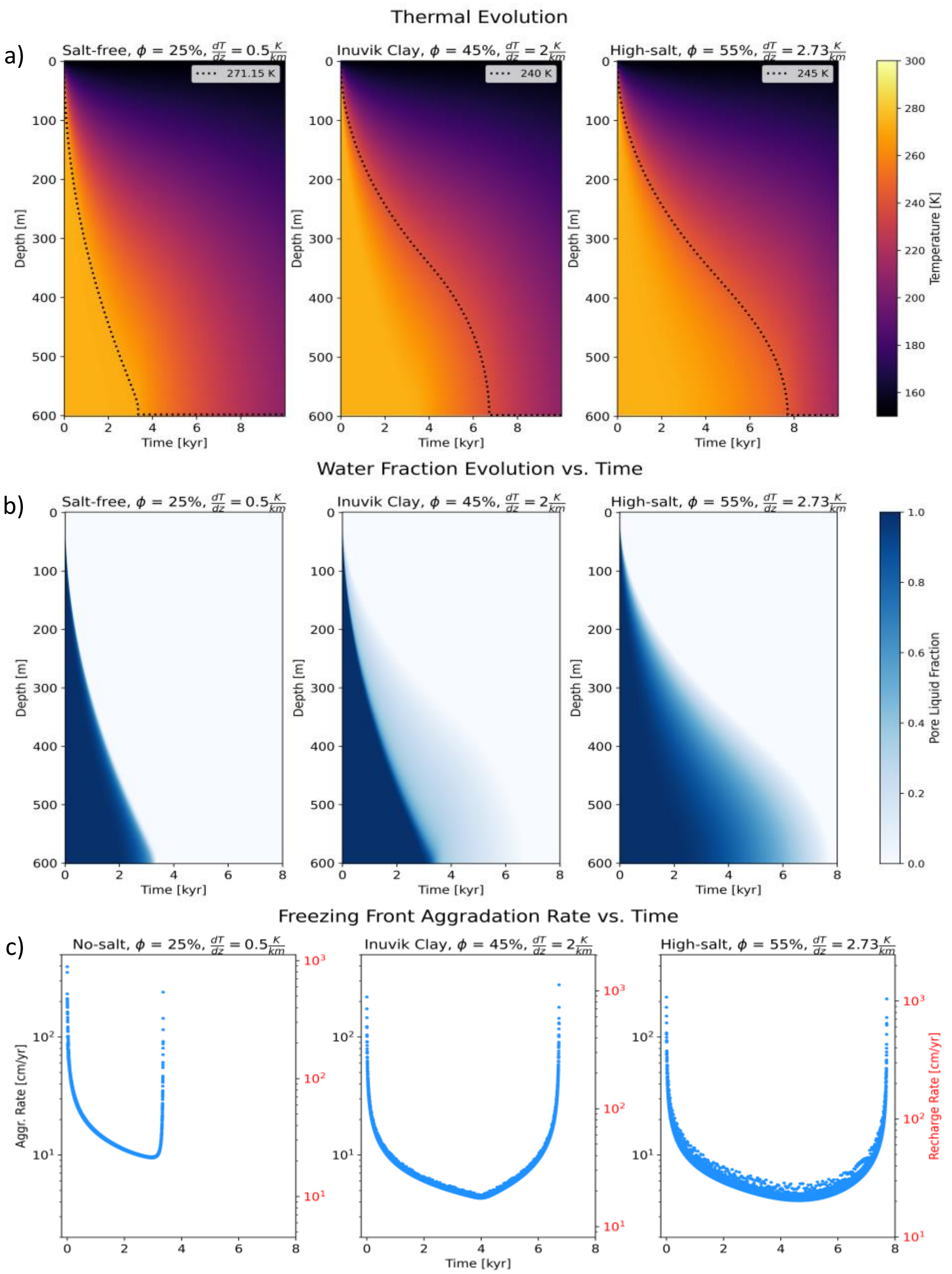

**Figure 6.** a) Thermal evolution, b) pore liquid fraction evolution, and c) freezing front aggradation rate and equivalent recharge rate (Eq. 4) for three thermal model runs. Total freeze out times range from ~3–8 kyr and are mostly modulated by the SFC. The timing and amount of pore water freezing dictates the amount of groundwater in the system that is usable by pingos on the surface.

Figure 6c shows freezing front aggradation rates and corresponding recharge rates per year while the impact melt freezes. The salt-free SFC case produced the greatest amount of recharge at the onset of freezing, causing deposition of large volumes of water in the near surface while the freezing front was shallow. However, recharge was not sustained for as long as the other cases. This case generated the minimum amount of water out of the entire suite of thermal model runs. The high-salt SFC, maximum porosity, maximum geothermal gradient case led to more consistent groundwater flow that was sustained for thousands of years. This case generated the maximum amount of water out of all thermal model runs. The median Inuvik Clay SFC case underwent similar freezing dynamics to the high-salt SFC case yet froze faster and generated less total water. See Supplement for the thermal, water fraction, and aggradation rate evolution of all 36 thermal model runs.

The groundwater flow model domain was comprised of columns of cells, each having a surface area of 10000 $m^2$. Figure 7 shows the cumulative total water delivery per column (100 m x 100 m) during freezing for each model run compared to the terrestrial case from Hornum et al. (2020). The timescales for the terrestrial and cerean model runs were comparable. Hornum et al.

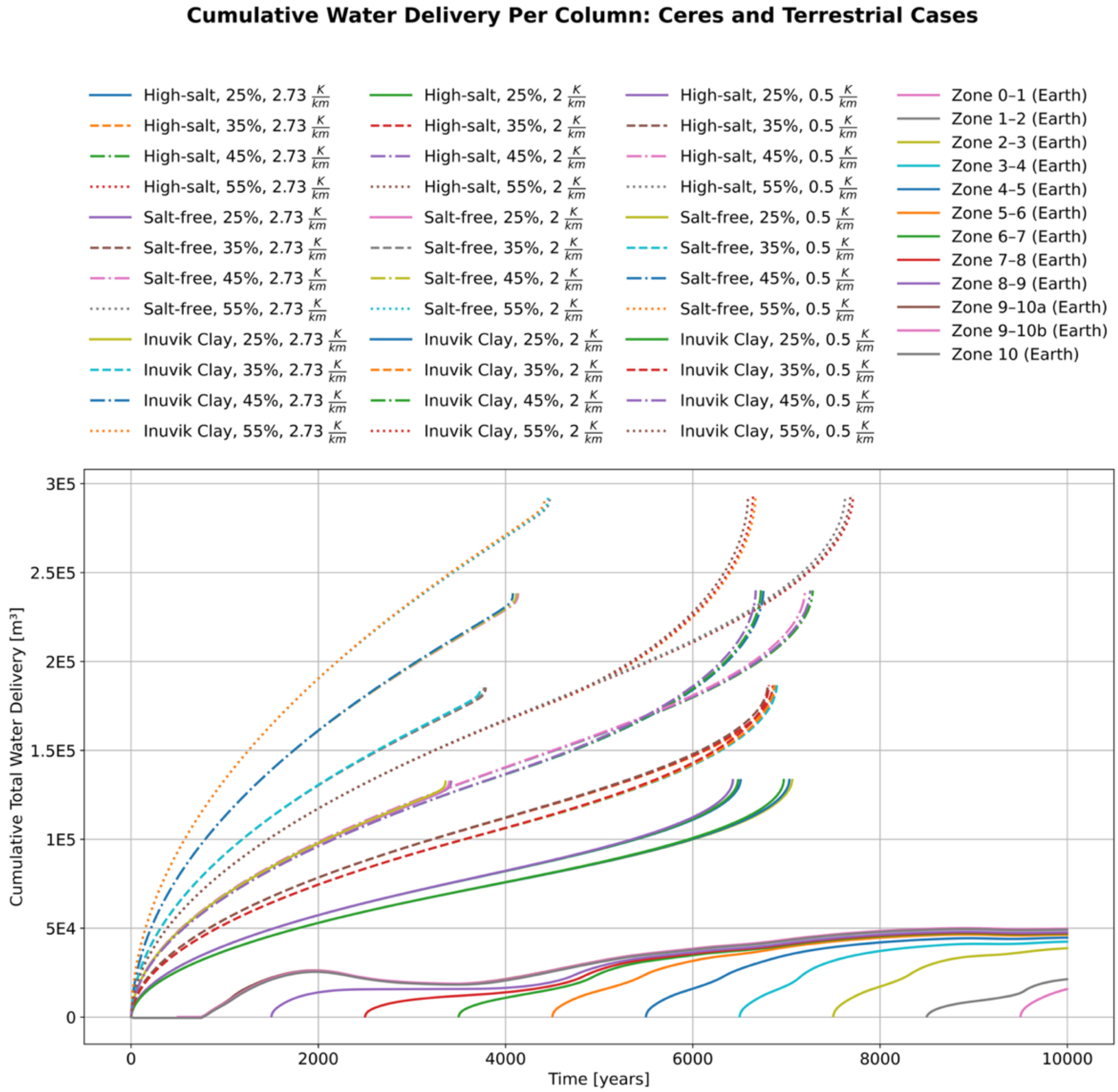

**Figure 7.** Cumulative total water delivery per column (recharge rate [m/yr] x 10000 $m^2$ x 1 year) as a function of time for Ceres and Earth cases. Earth case "zones" follow nomenclature in Hornum et al. (2020). Although the timescales are similar, the Ceres case yields an order of magnitude more water in the subsurface.

(2020) modeled pingo formation conditions in geologic units in Svalbard, Norway, which are a maximum of ~10 kyr old. Of our thermal model runs, the maximum elapsed time to total freezeout was ~8 kyr. Despite having similar timescales, the Ceres case yielded an order of magnitude more water than the terrestrial case. Interestingly, typical pingo candidate sizes on Ceres are an order of magnitude larger in diameter and height than Earth (Schmidt et al. 2020; Hughson et al. 2022). This size disparity may be explained by the difference in volumes of usable groundwater. On Earth, variable mean annual air temperatures and climate forcings over hundreds of years caused the freezing front to both aggrade and degrade. This led to final freezing front depths on the order of tens of meters to at most ~ 200 m (Hornum et al. 2020). A relatively shallow freezing front in tandem with low rock porosities (0.06–0.5) produced lower volumes of water. On Ceres, the lack of a variable climate prevented the freezing front from degrading. The freezing front advanced downward until the impact melt column was entirely frozen, expelling a maximum amount of pore water. However, not all groundwater was usable by pingos on the surface. Clays have relatively low permeabilities and high storage capacities when compared to other particulate materials. These attributes yielded very low hydraulic conductivities and rates of groundwater flow.

## 4.2 Groundwater Flow Simulations

We identified 20 pingo candidates in our study area (Figure 2). We used high-resolution DTMs of Occator Crater from Neesemann et al. (2025) to estimate their volumes, approximating each hill as a cone. Using geophysical data from Hughson et al. (2026) as a proxy for representative volumes of massive ice within terrestrial pingos, we assumed a range of ~10–90 volumetric percent. Our model tracked how much water was delivered to each pingo candidate during freezeout. If the water volume delivered to a hill was between 10–90% the hill's volume, we considered the threshold for pingo formation to be met. However, the exact volume of water necessary to create and sustain a terrestrial pingo is unknown. These bounds are loose constraints that likely underestimate the total volume of water that interacts with a pingo system during its lifetime (see section 5.2).

Figure 8 shows the discharge fractions for the hills in our study region for each of the 19 model scenarios. Notably, no model run yielded 20/20 hills as likely pingos. The 4 cases with the highest permeabilities (cases 16–19) produced an overabundance of water—in some cases, discharge volumes exceeded hill volumes by 100,000%. There are 3 categories of results: cases that did not generate enough water to create any pingos on the surface (cases 1 and 2), cases that generated excess water such that each hill received discharge > 90% its volume (cases 13–18), and cases that delivered water volumes optimal for pingo growth to a subset of hills on the surface (cases 3–12, 19). Although 11 of 20 model runs delivered optimal water volumes to surface hills, this proportion does not represent a probability of pingo occurrence. The simulations are not drawn from a statistical distribution but instead span a plausible parameter space. In each model run, no more than 7 out of 20 hills met the volume criterion (cases 5, 7, and 8). Other hills in the domain received deficient (e.g., case 3) or excess (e.g., case 12) water volumes. Case 19 was a special case. 19 out of 20 hills received excess water, but one hill received ~90% (Hill ID 15).

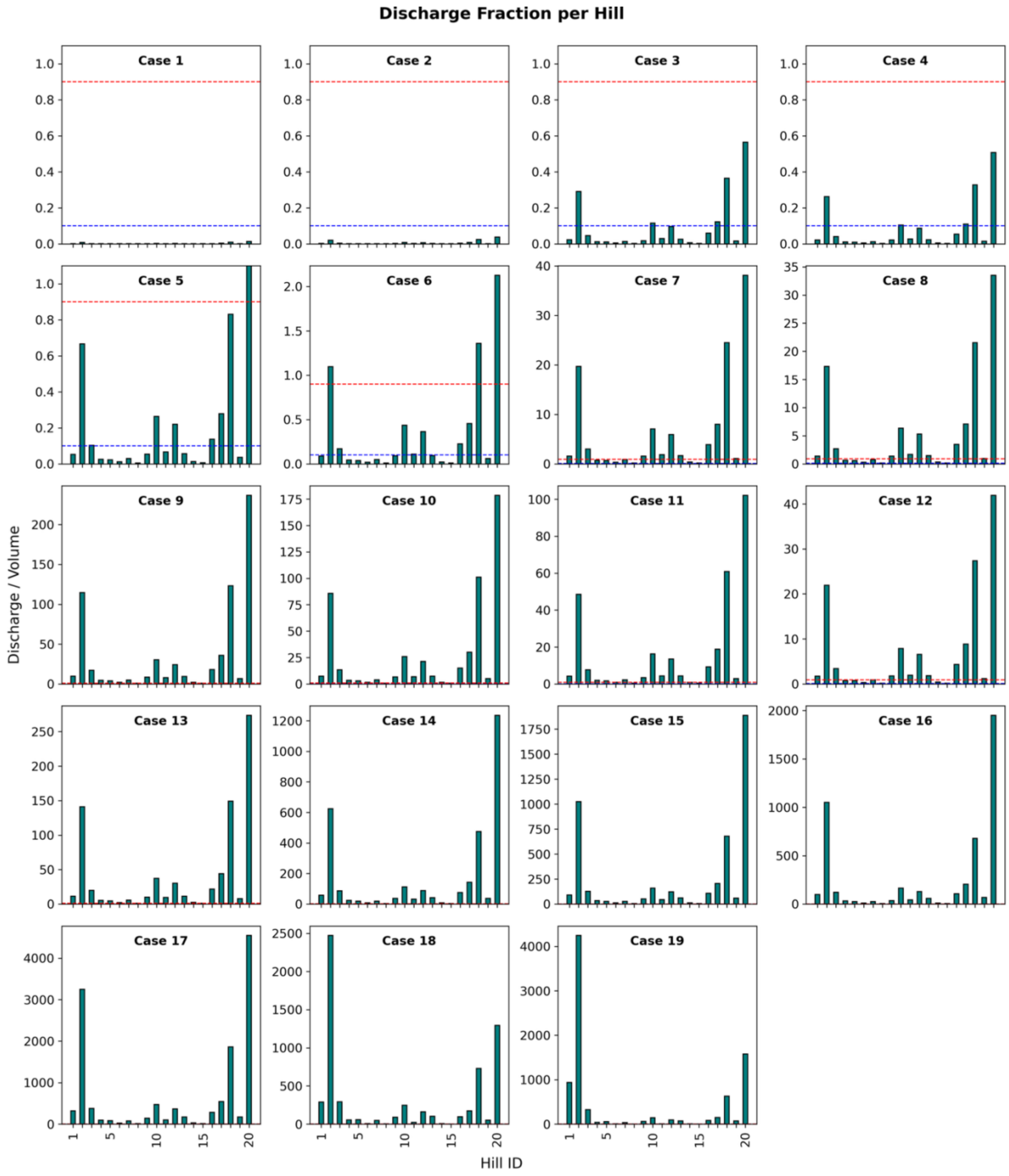

**Figure 8.** Total water volume discharge divided by hill volume for all 20 hills in all 19 cases. Hills with discharge fractions in the range of 10–90% are considered potential pingo candidates. These bounds are shown by dotted blue and red lines. Discharge fractions range from 0.000057 (Case 1) to 4548.19 (Case 17).

Cases 3 and 4 both delivered optimal water volumes to 5 hills on the surface. The other 15 hills received deficient volumes of water. Cases 5 and 6 delivered optimal water volumes to 7 and 6 hills on the surface, respectively. Some of the other hills received excess water, while others received deficient amounts. All other cases that delivered optimal water volumes to a subset of hills on the surface (Cases 7–12, 19) delivered excess water to all other hills in the domain. When permeability was greater than $10^{-19}$ m$^2$, excess water delivery was common. Notably, hills that received water volumes exceeding 90% of their own volume should not automatically be dismissed as pingo candidates. The amount of water interacting with pingo systems throughout their lifetimes is unconstrained, but it is certainly larger than the amount of ice within the pingo itself (see section 5.2). In most of the parameter space tested, groundwater discharge to hills easily surpassed the minimum volume required to form a pingo.

We tested permeabilities ranging from $10^{-20}$ to $10^{-12}$ m$^2$. High permeability ($10^{-12}$ to $10^{-14}$ m$^2$; consistent with sand, silt, loess, loam, and some layered clays) cases (16–19) exhibited different flow behavior from low permeability ($10^{-15}$ to $10^{-20}$ m$^2$; consistent with clays) cases (1–15). In the low-permeability regime (cases 1–15), all hills received an approximately equal amount of discharge within a given case (only ~20% variation in discharge volumes between hills was observed). The amount of discharge depends on our input parameters, primarily permeability and specific storage. Recharge and drain locations were applied to the uppermost active layer in our model, directly beneath the freezing front. Since this environment has an extremely low hydraulic conductivity (owing to the low permeability of clay and Ceres' reduced gravity), water could discharge through each drain before percolating downward or flowing laterally. We modeled each drain with the same conductance (Eq. 9), so they discharged water at the same rate. Another reason for the identical behavior of drains in our model runs was our assumed homogeneous subsurface. In the valley of Adventdalen, Svalbard examined by Hornum et al. (2020), stratigraphic information down to 1 km depths is known, and thus hydraulic conductivity varied laterally and with depth in their model. To model heat transfer and groundwater flow in cerean impact melts in more detail, we would need information on subsurface structure. Future geophysical landers could provide the subsurface information needed for more complex studies.

Drains behaved independently at higher permeabilities (~$10^{-14}$–$10^{-12}$ m$^2$). In this portion of the parameter space, topography allowed water to flow laterally along topographic gradients, and water could also percolate downward. Thus, drains received variable amounts of water in cases 16–19. Because the hills are all different sizes, we expected this behavior. However, these high permeability cases delivered extreme volumes of water to the hills. On the other end of the parameter space, low permeability cases generally yielded reasonable discharge to volume ratios, but the drains exhibited homogeneous behavior. We thus interpret these two flow regimes to be endmembers, with reality expected to fall between them. Since permeability actually varies vertically, laterally, and on different spatial scales, both endmember flow regimes may be realistic for different portions of the melt sheet in Occator Crater.

## 5 Discussion

### 5.1 Hydraulic Head and Fracture Potential

Hydraulic head is directly related to the excess pressure (i.e., fluid pressure – hydrostatic pressure) of a fluid and dictates groundwater flow. Groundwater flows along hydraulic head gradients, from high to low hydraulic head. Figure 9 shows the hydraulic head distribution throughout the model domain 1000 years into freezing for 3 endmember cases (see Supplement for all 19 cases). The layer shown is directly below the freezing front, where recharge is deposited and drains discharge water. The freezing front depth is dependent on material properties from the thermal model. We measured hydraulic head relative to the absolute minimum elevation in the model domain. If hydraulic head perfectly aligned with the topography of the impact melt itself, values would range from ~600–1700 m, because the impact melt is 600 m thick and spans 1150 m of relief. In some cases (e.g., 6), hydraulic head patterns mirror scarp locations in the impact melt deposit. Local elevation changes abruptly and affects groundwater flow in these regions.

In general, high permeability cases are consistent with observed topography and produce lower hydraulic heads than low permeability cases (e.g., case 16). Low permeability cases tend to have greater, more homogeneous, hydraulic head distributions. In the most extreme cases (e.g., case 6), hydraulic heads exceed 10 km. Hydraulic head is a proxy for pore pressure. Recharge adds pressure directly beneath the freezing front, and groundwater flows to dissipate the load. However, reduced permeability paired with Ceres' low gravity limits groundwater flow. In high permeability cases, groundwater flows vertically and along topographic gradients to distribute excess pressure. Groundwater flow is restricted in low permeability cases (e.g., case 1), causing pressure to increase homogeneously in the recharge layer directly below the freezing front.

In typical high permeability cases, water concentrates in the low-topography zone of the model domain (i.e., the northwest). Groundwater flows from high to low elevation but is stopped at the edge of the domain due to no-flow boundary conditions, artificially increasing hydraulic head. To investigate the impact this has on the results, we re-ran case 18 with a constant head boundary condition along the northwest rim of the melt deposit, allowing groundwater to flow out of the domain along topographic gradients that would better approximate real conditions at Ceres. After implementing the constant head boundary condition for case 18, hydraulic head values in the northwest were more consistent with the topography (see Supplement). However, this boundary condition minimally affected discharge through the hills, and thus the main conclusions about cerean pingo formation are unaffected. We also re-ran case 5 (low permeability regime) with the constant head boundary condition, which showed a similarly insignificant change to hill discharge and head distribution (see Supplement).

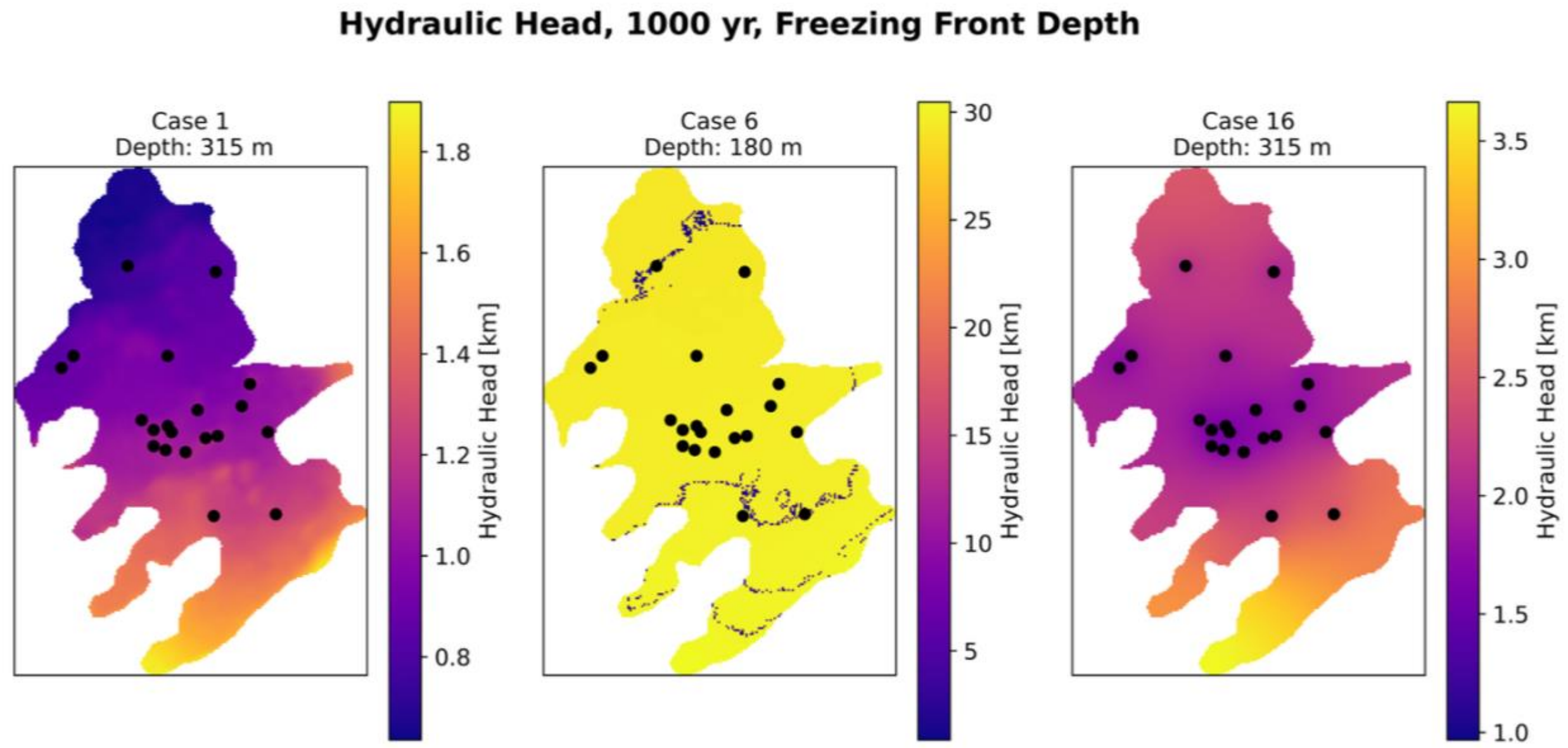

**Figure 9.** Hydraulic head distribution throughout the model domain 1000 yr into freezing for 3 endmember groundwater flow cases. Black dots are pingo candidate locations. The layer shown is directly below the freezing front depth, where recharge is applied and drains are located. Each case has a different freezing front depth set by thermal model parameters. Hydraulic head gradients directly affect the direction of groundwater flow.

Many pingo systems on Earth are topographically supported by sub-pingo water lenses under artesian conditions, so we expected hydraulic heads greater than surface elevations in our model (Mackay 1978). In all 19 model scenarios tested, pore water was pressurized enough for hydraulic head to lie above the ground surface. The range of vertical effective stresses (overburden stress – pore pressure) encountered under cerean conditions was -2.1E8 Pa (Case 15) to 5.7E4 Pa (Case 1). For effective stresses larger than -2 MPa, hydraulic fracturing is feasible. 15 out of 19 cases met the conditions for hydraulic fracturing, but the timing and location varied. This suggests fracturing of the freezing impact melt may have been common.

Figure 10 shows the time, depth, and locations of the onset of hydraulic fracturing for 3 endmember cases (see Supplement for all 19 cases). Case 3 first met the condition for hydraulic fracturing early into freezing in the south where elevation is highest. Case 14 met the condition for hydraulic fracturing 500 years into freezing, and the locations of hydraulic fracturing are modulated by the pingo “drains” actively discharging water. Case 18 did not meet the condition for hydraulic fracturing until water was confined to the bottom layer of the impact melt at the end of freezing. Since our model domain had no-flow boundary conditions, this pressurization was partially due to water being trapped by the top, bottom, and sides of the domain. However, unless the ground beneath the impact melt is heavily fractured, this boundary condition is physically realistic, as we would expect frozen crusts on the sides of the melt deposit to confine pore water. Portions of the floor of Occator are heavily fractured whereas others are likely to be relatively impervious. As discussed in section 3.3, portions of the melt sheet may be as thin as 100 m. Most cases meeting the condition for hydraulic fracturing do not do so until the freezing front is hundreds of meters deep. Thus, regions of impact melt on Ceres that lack visible fractures may signal a thinner local melt sheet or that subsurface fractures do not reach the surface.

Even though hydraulic fracturing may have occurred at depth, we cannot determine whether these fractures would reach the surface from these models. Fracture theory states that hydraulic fractures open perpendicular to the lowest principal stress, which, in our case, is oriented in the vertical direction (because pore pressure relieves lithostatic pressure in the vertical direction; Behrmann 1991). Thus, we expect hydraulic fractures to initially open horizontally and propagate. However, fracturing from high-strength and high-stress zones to lower-stress zones promotes vertical fracture propagation, which may be relevant in this context (Liao et al. 2024). If hydraulic fractures propagated to the surface, they may have provided pathways for surface extrusion of brines without relying on a deep crustal brine reservoir. Nevertheless, groundwater flow models suggest that hydraulic fracturing may be a common process that influences pressure dissipation and groundwater flow within shallow impact melts. The inclusion of fractures in our model domain was outside the scope of this investigation. However, the nature of hydraulic fracturing generated by impact melt freezing and the effect of fracturing on pressure relief and groundwater flow on Ceres would be a rich subject for future numerical studies.

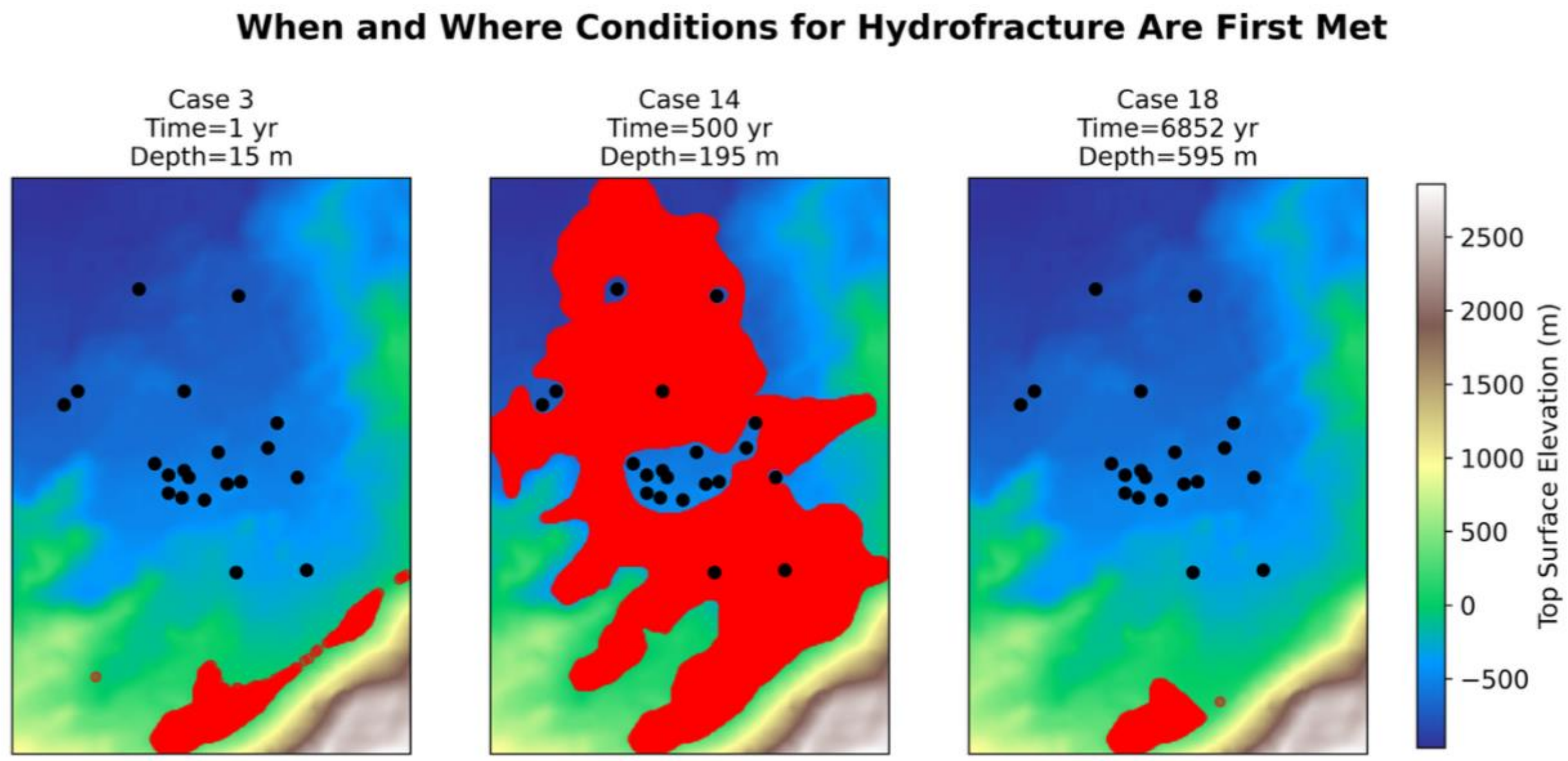

**Figure 10.** Model cells that met the conditions for hydraulic fracturing (pore pressure > overburden stress + 2 MPa) for 3 endmembers cases (red). The colormap is surface topography and the black dots are pingo candidate locations. Hydraulic fracturing may open additional pathways for groundwater flow.

## 5.2 Implications for Pingo Formation

Permeability is not the sole factor influencing groundwater flow, as illustrated by the variability in hydraulic head and discharge volumes seen within the high and low permeability regimes. Impact melts with greater porosities contain more pore water that can be expelled. Porosity is closely tied to permeability, which is an important factor for hydraulic conductivity. We expect large porosities and permeabilities to yield more groundwater flow, which is essential for pingo growth but can also supply excess water in aquifers with a low specific storage. Specific storage is related to an aquifer’s capacity to release groundwater for a given change in head. Greater specific storage leads to more water trapped by the clay minerals, whereas lower specific storage releases water more readily. Compared to other aquifer materials, clay is characterized by fine-grained particles and high compressibility values that lead to a relatively high specific storage (Freeze and Cherry, 1979). Larger geothermal gradients prevent total freezeout of the impact melt and allow groundwater to flow for longer timescales. Thus, we also expect larger geothermal gradients to be more conducive to sustained groundwater flow. A geothermal gradient of 0.5 K/km (Bowling et al. 2019) does not preserve liquid water long enough to form any pingos on the surface, unless it is paired with an impact melt with a minimum specific storage. A geothermal gradient of 2.73 K/km (Raymond et al. 2020) keeps liquid in the impact melt unfrozen for the longest time but can deliver excessive water to the surface if paired with a high porosity/permeability or low specific storage. A median geothermal gradient of 2 K/km (Hesse & Castillo-Rogez 2019) balances liquid water longevity and drain discharge to create favorable conditions for pingo growth when paired with median porosities/permeabilities and specific storage values. Perhaps surprisingly, our choice of SFC does not strongly affect the volume of water delivery. Although the presence of salt depresses the freezing point of water and may help sustain groundwater flow, ultimately the physical properties of the impact melt (porosity/permeability and specific storage) are more important constraints on the plausibility of pingo formation.

Figure 11 shows the hills that meet our discharge volume criteria for pingo growth for all groundwater flow simulations (i.e., discharge volumes 10–90% of hill volumes). The cases that delivered optimal discharge to surface hills were cases 3–12 and 19. The two pingo candidates that meet the criteria most often in our model runs are hills 15 and 8. These hills are morphologically the most compelling pingo candidates in our model domain, due to their ruptured surfaces, circular planform geometries, and proximity to a salt deposit (shown in Figure 1), which is direct evidence of brine surface extrusion (Scully et al. 2018). There are 5 cases in which hills 15 and 8 both meet the criteria for pingo formation (cases 7, 8, 10, 11, and 12). These cases have 45% or 55% porosities, $10^{-16}$–$10^{-15}$ $m^2$ permeabilities, and geothermal gradients of 2 or 2.73 K/km. All 3 SFCs and aquifer compressibility values are represented. These values are intermediate in our parameter space and consistent with median-porosity clays.

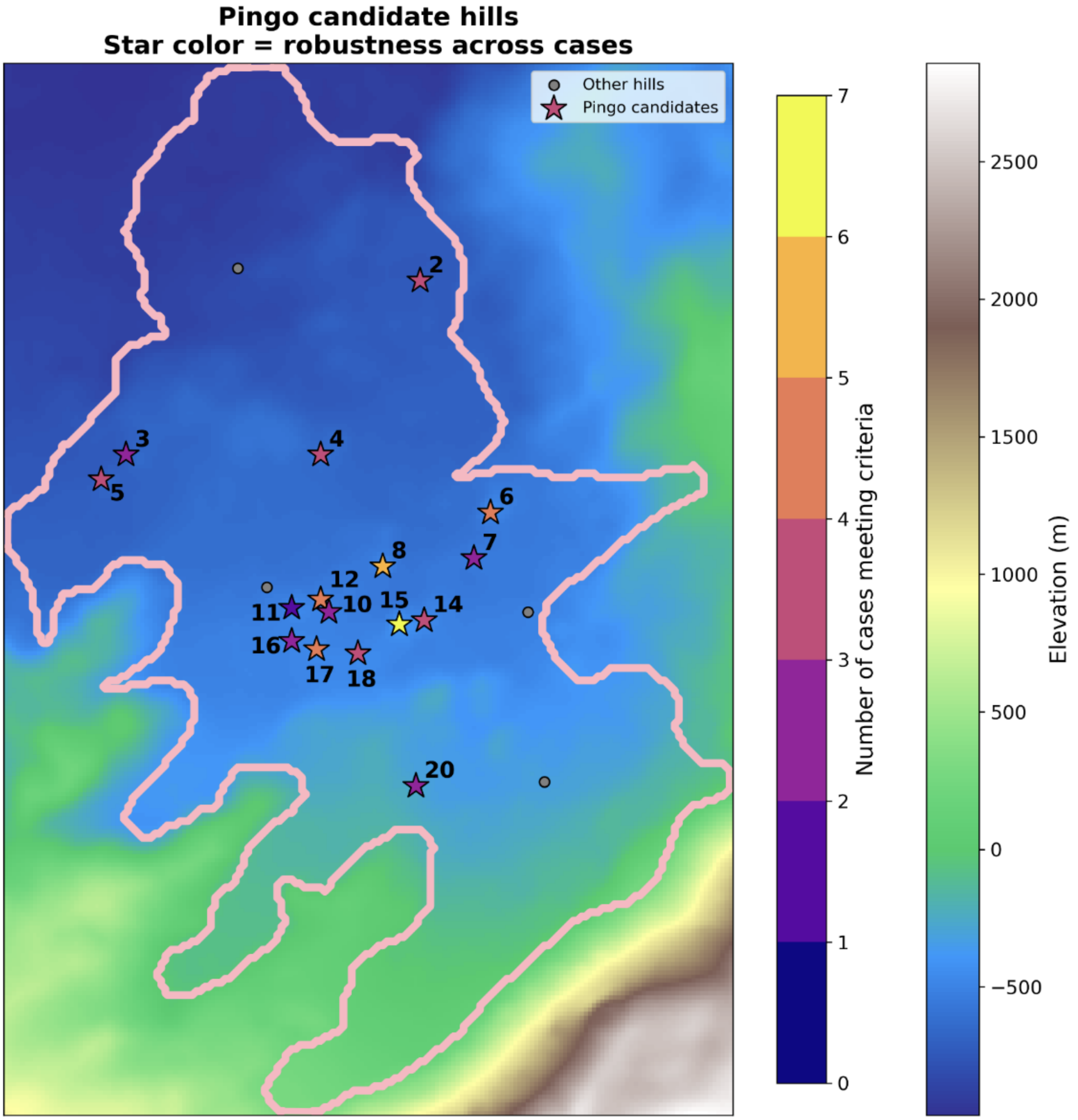

**Figure 11.** Hills in our model domain that meet the pingo discharge fraction criteria (stars), colored by the number of cases in which they meet the criteria. The colormap is surface topography. The two hills that meet the criteria most often (Hill IDs 8 and 15) are also the most morphologically compelling pingo candidates in the model domain.

Our defined ice volume bounds of 10–90% for cerean pingo candidates are loosely constrained from radargrams of terrestrial pingos (Hughson et al. 2026). However, it is unknown how much water is delivered to pingos throughout their lifetimes, how much is lost through seasonal springs, or how efficiently water delivered to a pingo is incorporated into the ice core. Terrestrial pingos grow in saturated taliks that provide groundwater to accrete onto pingo ice cores. Even though this water does not count toward the volume of the pingo, it is still an essential part of terrestrial pingo formation and evolutionary processes; thus, some cases where discharge volumes exceed the 90% threshold may be relevant for pingo formation as well. We chose a threshold based off estimates derived from terrestrial data and explore what would happen if we chose another one speculatively.

In an attempt to put more accurate volumetric bounds on the amount of water needed to form a pingo, we used typical hydrostatic pingo systems in the Mackenzie Delta region as an example. These pingos form in the centers of drained thermokarst lakes. However, only a fraction of the initial lake's volume goes toward the formation of an individual pingo. The rest of the talik supplies groundwater to growing pingos for hundreds to thousands of years (in addition to recharge from precipitation and discharge through springs). Ibyuk pingo, for example, has a volume of ~$1.1 \times 10^6$ $m^3$. It grew in a drained thermokarst lake with an approximate surface area of ~$3 \times 10^6$ $m^2$. If we assume the talik depth was 30 m (Mackay 1986) and approximate the subsurface as half an ellipsoid, the total talik volume was $8 \times 10^7$ $m^3$. Assuming 30% porosity (typical of lacustrine sediments), the saturated talik housed $2.4 \times 10^7$ $m^3$ of groundwater. This is 22 times greater than the volume of Ibyuk. However, as the talik froze, pore water expulsion created pressure gradients that forced expelled water toward the center of the drained lake. Only 9% of pore water may be expelled from the pore space, so the most realistic volume of water in the talik that contributed to the formation of Ibyuk was ~2 times the volume of the pingo. In our simulations, we did not convert water discharge volumes into ice accumulation volumes, because not all water used by a pingo system ends up as ice inside the pingo itself. However, if we were to assume some of the discharged water freezes and expands volumetrically, the lower bound would be less than 10%. By these approximations, our initial bounds of 10–90% were not unreasonable, but they likely slightly overestimate the minimum water volume required for pingo formation and underestimate the maximum water volume that may be involved in some settings.

What we refer to as "excess" water delivery (>90% of hill volume) was common in our simulations. There were two groups of model scenarios that delivered excess water to all hills in the study area. Cases 13–15 had high porosities (55%), median permeabilities ($10^{-15}$ $m^2$), high geothermal gradients (2.73 K/km), and low specific storages ($10^{-5}$ or $10^{-6}$ $m^{-1}$). Cases 16–19 were high permeability endmembers ($10^{-12}$–$10^{-14}$ $m^2$) that spanned all porosities, geothermal gradients, and SFCs. Cases 16–19 also had low specific storages ($10^{-5}$ or $10^{-6}$ $m^{-1}$). Among the model scenarios that delivered optimal water volumes (10%–90% of hill volume) to a subset of hills in the study region (cases 3–12, 19), other hills in the domain experienced excess water delivery—i.e., discharge volumes 10s to 100s times greater than the corresponding hill volume.

If we re-define the threshold for "excess" water delivery from 90% to 200% of hill volume, Cases 7, 8 and 12 each yield 12 pingo candidates with "optimal" growth conditions. Additionally, Cases 13 and 18 join the group of scenarios that supply "optimal" water volumes to at least one hill. Extending the "excess" threshold to 200% allows every hill in the domain to fall in the "optimal" range for pingo formation in at least one scenario we explored, underscoring the fact that oversupply of water is a more common prediction of our simulations than insufficient water supply.

We cannot assume that "excess" water scenarios (>90% or >200%) preclude pingo formation, or that they are unphysical depictions of impact melt freezing processes in Occator. At some terrestrial locations, groundwater from sub-pingo water lenses flows to the surface through pingo dilation fractures, forming springs. Pingo spring discharge rates are scarce in the literature (Yoshikawa & Harada 1995; Liestøl 1996), and the total amount of water lost via spring flow during a terrestrial pingo's lifetime is unknown. An analogous process may operate on Ceres, with excess groundwater flowing to the surface through dilation cracks or tensile fractures, then rapidly sublimating. In this scenario, we might expect short, sinuous channels and/or evaporite deposits to form at spring locations. However, these features may not be resolvable in current Dawn imagery (3 m/pix) of our study region and they could have been removed by surface weathering and impact gardening over millions of years. High discharge scenarios may be responsible for other prominent features in Occator Crater, such as Cerealia and Vinalia faculae (see Figure 2). These large salt deposits were previously hypothesized to be sourced from a deep crustal brine layer. However, our results show that abundant groundwater can be delivered to the near surface from impact melt freezing itself, perhaps providing an alternate formation mechanism for the faculae if our assumptions about the initial volatile fraction of the melt are accurate. This possibility should be the focus of future theoretical investigations.

Alternatively, GRaND-and VIR data-derived maps of Occator Crater showed hydrogen concentrations spanning 16–19 and 16–17 wt.% eq. $H_2O$, respectively (Prettyman et al. 2021). Although lobate impact melt deposits may retain significantly more, it is possible that the impact melt is less volumetrically water-rich than we accounted for in our models (our minimum porosity was 25%). Since many of our high-porosity, high-permeability cases delivered excess water to hills on the surface, considering our results and morphological observations of the surface together may suggest that the impact melt is less water-rich than minimum water volume scenarios in our parameter space. Another explanation for the lack of abundant, obvious pingo-like features and hydrothermal activity in our model domain may be thinner regions of impact melt that locally reduce water volumes by a factor of 2–6, or by the effects of high salt concentration and low temperature on viscosity that were neglected in our model (section 3.3).

### 5.3 Cerean Pingos in a Solar System Context

Our results indicate pingo formation is plausible on Ceres for a significant portion of the parameter space we explored. The formation process is different from the terrestrial case. On Earth, pingos grow or subside annually via seasonal freeze-thaw cycles that allow water to enter or leave the pingo system. Ceres lacks these cycles, and thus potential pingos have one chance to form as the impact melt freezes. However, the total timescale for growth is comparable in both settings (Figure 7). Pingos on Earth reach the end of their lifecycles when their ice cores become exposed to the atmosphere and thaw, creating slumping and eventual collapse (Mackay 1998). This may be triggered by warm temperature anomalies or by overburden fracturing, but we do not expect significant changes to thermal environments on Ceres. Several pingo candidates in our study region are similarly fractured but not collapsed. This suggests that ice has not had enough time to fully sublimate, as it is protected by a centimeters–meters thick overburden. With the awareness that buried ice at a temperature of 145 K is stable on Ceres, ice buried a few meters beneath dusty surfaces may be stable for billions of years for a grain size ~0.1 mm (Schorghofer 2008; Hayne & Aharonson 2015). The average grain size of the surface layer of floor units in Occator Crater derived from NIR spectroscopy is ~0.1 mm, although this measurement carries significant uncertainty (Raponi et al. 2019). Away from the poles, where the ice table is deeper, buried ice is also protected from impact gardening-driven ice loss (Schorghofer 2019). Ibyuk pingo on Earth has an overburden of ~15 m (Müller 1962). If we assume similar overburden thicknesses on Ceres, this buried ice is plausibly stable on timescales of billions of years.

Ceres also differs from Earth in terms of subsurface pressure regime. We have shown that on Ceres, pore pressures can exceed both the overburden stress and tensile strength of impact melt. Vertical effective stress is always positive in the terrestrial simulations of Hornum et al. 2020, ensuring that conditions for hydraulic fracture are never met in Svalbard. Impact melts in Occator Crater exist in an entirely different—and more extreme—stress regime. Constant top-down freezing creates an impermeable lid that pressurizes the pore water beneath it. Additionally, Ceres' low gravity leads to lower overburden stresses than encountered on Earth and low hydraulic conductivities (Eq. 7). Both inflate pore pressures and prevent groundwater from effectively dissipating pressure, allowing pore pressure to easily overcome overburden stress and create hydraulic fractures.

Ceres is not the only planet that hosts fractured domes and hills atop impact melt deposits in crater floors. On the Moon, Tycho crater hosts similar features, although the Moon is extremely water-poor, suggesting an impact melt-driven formation process for small fractured hills (Dhingra et al. 2017). Pingo candidates have also been identified in the Martian mid-latitudes (Dundas et al. 2008). Some of the fractured martian hills occur on crater floors and others are near gullies, but the case for interpreting each of these features as pingos is debated (Dundas & McEwen, 2010). The case for pingos on Ceres is more plausible, especially in impact craters. Typical large impacts at Ceres release energy sufficient to melt subsurface ice but not rock. Impacts that produce comparably sized craters on Mars produce much higher temperatures that typically devolatilize the target material (Tornabene et al. 2012; Sizemore et al. 2017). Several studies have investigated subsurface water flow and excess ice generation on Mars (e.g., Mellon & Phillips 2001; Lacelle et al. 2008; Fisher & Lacelle 2014; Marra et al. 2014; Sizemore et al., 2015; Fisher et al. 2022), but a dedicated pingo formation model has yet to be developed to investigate the plausibility of pingos on Mars. Even in the absence of pingo-like features, thermal- and pressure-driven groundwater flow is likely a ubiquitous process in the shallow subsurface environments of several bodies in the solar system: including ice-rich asteroids and icy moons like Europa, Enceladus, and Titan. Hydrologic processes impact the development and transience of habitable subsurface environments, planetary landscape evolution, and in situ resource utilization.

## 6 Conclusions

Our coupled thermal and groundwater flow models demonstrate that impact melt freezing produces pressure gradients sufficient to drive groundwater flow in the shallow subsurface (< 1 km depths) and that pingo formation is plausible in the cerean post-impact environment. We adapted a terrestrial pingo formation model to the unique physical and thermal environment of Occator Crater, testing a comprehensive parameter space that includes porosity, permeability, specific storage, geothermal gradient, and soil freezing curves (SFCs) for different salt concentrations and particle geometries. There is an intermediate regime in our parameter space that yields optimal water delivery consistent with observed, pingo-like hill morphologies. The cases that delivered favorable discharge volumes to the most morphologically compelling hills in Occator Crater have porosities of 45–55%, permeabilities of $10^{-16}$–$10^{-15}$ $m^2$, and geothermal gradients of 2 or 2.73 K/km. All specific storage values and soil freezing curves are represented. These values are consistent with median-porosity clays, as would be expected for an impact melt slurry on a low-gravity planet. In several cases, pore pressures exceed the overburden stress, indicating that hydraulic fracturing may facilitate fluid transport, generate permeability heterogeneities, and promote ice lens formation. Our results suggest that groundwater flow could produce accessible near-surface ice and potentially localized brine extrusion at Occator Crater.

Our results have broader implications for the study of near-surface hydrology on volatile-rich planetary bodies. The model developed here provides a template for investigating similar morphologies on Mars and icy moons. Future theoretical work should include refinement of subsurface properties for cerean materials and fracture mechanics. Moreover, the presence of near-surface massive ice and potential localized brine extrusion formed through this process may present compelling targets for future exploration, particularly for missions focused on astrobiology and in situ resource utilization. Future missions could test these predictions through high-resolution (< 1 m/pix) imaging and spectral observations capable of identifying surface expressions of brine extrusion. Subsurface geophysical measurements, such as ground-penetrating radar, could further constrain stratigraphy, characterize bedrock topography, and detect massive ice or fracture networks within the impact melt. Returned samples could potentially inform grain size, porosity, permeability, density, and thermodynamic properties (e.g., thermal conductivity) (e.g., Castillo-Rogez et al., 2022). These data products would facilitate more detailed and accurate models, allow us to directly test the pingo hypothesis, and advance our understanding of cryohydrology in the solar system.

**Acknowledgements**

A. R. K. was supported by the NASA FINESST grant NNH22ZDA001N. H. G. S. and B. E. S. were supported by NASA grant 80NSSC22K1431. We thank R. Winston and S. Boyce at USGS for their help with MODFLOW code.

**Availability Statement**

The thermal model python script, thermal model and MODFLOW input and output files, shapefiles used for the groundwater flow model, and scripts used for data analysis and visualization are publicly available through Zenodo (Kubas et al. 2026).

**Conflict of Interest**
The authors declare there are no conflicts of interest for this manuscript.

**References**
Andersland, O. B., & Ladanyi, B. (2003). *Frozen Ground Engineering*. John Wiley & Sons.

Behrmann, J. H. (1991). Conditions for hydrofracture and the fluid permeability of accretionary wedges. *Earth and Planetary Science Letters*, *107*(3), 550–558. https://doi.org/10.1016/0012-821X(91)90100-V

Bergman, T. L. (2011). *Fundamentals of Heat and Mass Transfer*. John Wiley & Sons.

Bland, M. T., Raymond, C. A., Schenk, P. M., Fu, R. R., Kneissl, T., Pasckert, J. H., Hiesinger, H., Preusker, F., Park, R. S., Marchi, S., King, S. D., Castillo-Rogez, J. C., & Russell, C. T. (2016). Composition and structure of the shallow subsurface of Ceres revealed by crater morphology. *Nature Geoscience*, *9*(7), 538–542. https://doi.org/10.1038/ngeo2743

Bowling, T. J., Ciesla, F. J., Davison, T. M., Scully, J. E. C., Castillo-Rogez, J. C., Marchi, S., & Johnson, B. C. (2019). Post-impact thermal structure and cooling timescales of Occator crater on asteroid 1 Ceres. *Icarus, Occator Crater on Ceres*, *320*, 110–118. https://doi.org/10.1016/j.icarus.2018.08.028

Castillo-Rogez, J. C., Hesse, M. A., Formisano, M., Sizemore, H., Bland, M., Ermakov, A. I., & Fu, R. R. (2019). Conditions for the Long-Term Preservation of a Deep Brine Reservoir in Ceres. *Geophysical Research Letters*, *46*(4), 1963–1972. https://doi.org/10.1029/2018GL081473

Castillo-Rogez, J. C., & McCord, T. B. (2010). Ceres' evolution and present state constrained by shape data. *Icarus*, *205*(2), 443–459. https://doi.org/10.1016/j.icarus.2009.04.008

Castillo-Rogez, J., Neveu, M., McSween, H. Y., Fu, R. R., Toplis, M. J., & Prettyman, T. (2018). Insights into Ceres's evolution from surface composition. *Meteoritics & Planetary Science*, *53*(9), 1820–1843. https://doi.org/10.1111/maps.13181

Castillo-Rogez, J., Brophy, J., Miller, K., Sori, M., Scully, J., Quick, L., Grimm, R., Zolensky, M., Bland, M., Buczkowski, D., Raymond, C., Hendrix, A., Prettyman, T., Sekine, Y., Titus, T., Williams, D., Backes, P., Barge, L., Ermakov, A., Galassi, A., Moreland, S., & Zacny, K. (2022). Concepts for the Future Exploration of Dwarf Planet Ceres' Habitability. *The Planetary Science Journal*, *3*(2), 41. https://doi.org/10.3847/PSJ/ac34ee

Chen, J., Mei, S., Irizarry, J. T., & Rempel, A. W. (2020). A Monte Carlo Approach to Approximating the Effects of Pore Geometry on the Phase Behavior of Soil Freezing. *Journal of Advances in Modeling Earth Systems*, *12*(10). https://doi.org/10.1029/2020MS002117

Chilton, H. T., Schmidt, B. E., Duarte, K., Ferrier, K. L., Hughson, K. H. G., Scully, J. E. C., Wray, J. J., Sizemore, H. G., Nathues, A., Platz, T., Schorghofer, N., Schenk, P. M., Landis, M. E., Bland, M., Byrne, S., Russell, C. T. R., & Raymond, C. A. (2019). Landslides on Ceres: Inferences Into Ice

Content and Layering in the Upper Crust. *Journal of Geophysical Research: Planets*, *124*(6), 1512–1524. https://doi.org/10.1029/2018JE005634

Dash, J. G., Rempel, A. W., & Wettlaufer, J. S. (2006). The physics of premelted ice and its geophysical consequences. *Reviews of Modern Physics*, *78*(3), 695–741. https://doi.org/10.1103/RevModPhys.78.695

Delpech, L.-M., Tveit, A. T., Hodson, A. J., Hand, K. P., Kalentichenko, D. (2024). Chemolithoautotrophic bacteria flourish at dark water–ice interfaces of an emerged Arctic cold seep. *The ISME Journal, 18(1),* https://doi.org/10.1093/ismejo/wrae170

Dhingra, D., Head, J. W., & Pieters, C. M. (2017). Geological mapping of impact melt deposits at lunar complex craters Jackson and Tycho: Morphologic and topographic diversity and relation to the cratering process. *Icarus, Lunar Reconnaissance Orbiter - Part II, 283*, 268–281. https://doi.org/10.1016/j.icarus.2016.05.004

Douglas, T. A., & Mellon, M. T. (2019). Sublimation of terrestrial permafrost and the implications for ice-loss processes on Mars. *Nature Communications*, *10*(1), 1716. https://doi.org/10.1038/s41467-019-09410-8

Dundas, C. M., & McEwen, A. S. (2010). An assessment of evidence for pingos on Mars using HiRISE. *Icarus, MRO/HiRISE Studies of Mars*, *205*(1), 244–258. https://doi.org/10.1016/j.icarus.2009.02.020

Dundas, C. M., Mellon, M. T., McEwen, A. S., Lefort, A., Keszthelyi, L. P., & Thomas, N. (2008). HiRISE observations of fractured mounds: Possible Martian pingos. *Geophysical Research Letters*, *35*(4). https://doi.org/10.1029/2007GL031798

Ermakov, A. I., Fu, R. R., Castillo-Rogez, J. C., Raymond, C. A., Park, R. S., Preusker, F., Russell, C. T., Smith, D. E., & Zuber, M. T. (2017). Constraints on Ceres' Internal Structure and Evolution From Its Shape and Gravity Measured by the Dawn Spacecraft. *Journal of Geophysical Research: Planets*, *122*(11), 2267–2293. https://doi.org/10.1002/2017JE005302

Ermakov, A. I., Mazarico, E., Schröder, S. E., Carsenty, U., Schorghofer, N., Preusker, F., Raymond, C. A., Russell, C. T., & Zuber, M. T. (2017), Ceres's obliquity history and its implications for the permanently shadowed regions, *Geophys. Res. Lett.*, 44, 2652–2661, doi:10.1002/2016GL072250.

Fetter, C. W. (2018). *Applied Hydrogeology: Fourth Edition*. Waveland Press.

Fisher, D. A., & Lacelle, D. (2014). A model for co-isotopic signatures of evolving ground ice in the cold dry environments of Earth and Mars. *Icarus*, *243*, 454–470. https://doi.org/10.1016/j.icarus.2014.08.009

Fisher, D., Lacelle, D., & Pollard, W. (2022). A possible perchlorate-enabled mechanism for forming thick near surface excess ice layers; in the Amazonian regolith of Mars. *Icarus*, *387*, 115198. https://doi.org/10.1016/j.icarus.2022.115198

Freeze, R. A., & Cherry, J. A. (1979). *Groundwater*. Prentice-Hall.

Govaerts, J., Beerten, K., & ten Veen, J. (2016). Weichselian permafrost depth in the Netherlands: A comprehensive uncertainty and sensitivity analysis. *The Cryosphere*, *10*(6), 2907–2922. https://doi.org/10.5194/tc-10-2907-2016

Grindrod, P. M., Fortes, A. D., Nimmo, F., Feltham, D. L., Brodholt, J. P., & Vočadlo, L. (2008). The long-term stability of a possible aqueous ammonium sulfate ocean inside Titan. *Icarus*, *197*(1), 137–151. https://doi.org/10.1016/j.icarus.2008.04.006

Hayne, P. O., & Aharonson, O. (2015), Thermal stability of ice on Ceres with rough topography, *J. Geophys. Res. Planets, 120,* 1567–1584, doi:10.1002/2015JE004887.

Hesse, M. A., & Castillo-Rogez, J. C. (2019). Thermal evolution of the impact-induced cryomagma chamber beneath Occator Crater on Ceres. Geophysical Research Letters, 46, 1213–1221. https://doi.org/10.1029/2018GL080327

Hiesinger, H., Marchi, S., Schmedemann, N., Schenk, P., Pasckert, J. H., Neesemann, A., O'Brien, D. P., Kneissl, T., et al. (2016). Cratering on Ceres: Implications for its crust and evolution. *Science, 353*(6303), doi:10.1126/science.aaf4759

Hornum, M. T., Hodson, A. J., Jessen, S., Bense, V., & Senger, K. (2020). Numerical modelling of permafrost spring discharge and open-system pingo formation induced by basal permafrost aggradation. *The Cryosphere*, *14*(12), 4627–4651. https://doi.org/10.5194/tc-14-4627-2020

Hughson, K. H. G., Russell, C. T., Schmidt, B. E., Chilton, H. T., Sizemore, H., Schenk, P. M., & Raymond, C. A. (2019). Fluidized appearing ejecta on Ceres: Implications for the mechanical properties, frictional properties, and composition of its shallow subsurface. *Journal of Geophysical Research: Planets, 124,* 1819–1839. https://doi.org/10.1029/2018JE005666

Hughson, K. H. G., Russell, C. T., Williams, D. A., Buczkowski, D. L., Mest, S. C., Pasckert, J. H., Scully, J. E. C., Combe, J.-P., Platz, T., Ruesch, O., Preusker, F., Jaumann, R., Nass, A., Roatsch, T., Nathues, A., Schaefer, M., Schmidt, B. E., Chilton, H. T., Ermakov, A., … Raymond, C. A. (2018). The Ac-5 (Fejokoo) quadrangle of Ceres: Geologic map and geomorphological evidence for ground ice mediated surface processes. *Icarus, The Geologic Mapping of Ceres*, *316*, 63–83. https://doi.org/10.1016/j.icarus.2017.09.035

Hughson, K. H. G., Schmidt, B. E., Udell Lopez, K. T., Sizemore, H. G., Schenk, P. M., Scully, J. E. C., Raymond, C. A., & Russell, C. T. (2022). Comparative morphometric analysis suggests ice-cored pingo-shaped landforms on the dwarf planet Ceres. *Geology*, *50*(4), 522–527. https://doi.org/10.1130/G49321.1

Hughson, K. H. G., Schmidt, B. E., Siegfried, M. R., Bradford, J. H., Sizemore, H. G., Swidinsky, A., Kubas, A. R., Routt, A. A., Mullen, A. D., Michaelides, R. J., Hills, B. H., Follingstad, V. R., Quartini, E., Lein, D. J., Noh, K., Cockney, R., & Gruben, G. (2026). Diverse Structure of Pingos in Arctic Permafrost Reveal Complex Trajectories of Ice Growth. *Science Advances,* Manuscript submitted for publication.

Jakosky, B. M., Nealson, K. H., Bakermans, C., Ley, R. E., & Mellon, M. T. (2003). Subfreezing Activity of Microorganisms and the Potential Habitability of Mars' Polar Regions. *Astrobiology*, *3*(2), 343–350. https://doi.org/10.1089/153110703769016433

King, S. D., Castillo-Rogez, J. C., Toplis, M. J., Bland, M. T., Raymond, C. A., & Russell, C. T. (2018). Ceres internal structure from geophysical constraints. *Meteoritics & Planetary Science*, *53*(9), 1999–2007. https://doi.org/10.1111/maps.13063

Kubas, A. R., Sizemore, H., & Schmidt, B. E. (2026). Data and Scripts from: Cryohydrologic Flow and Pingo Formation Within Freezing Impact Melts in Occator Crater, Ceres [Data set]. Zenodo. https://doi.org/10.5281/zenodo.20312337

Lacelle, D., Fisher, D., Clark, I. D., & Berinstain, A. (2008). Distinguishing between vapor- and liquid-formed ground ice in the northern martian regolith and potential for biosignatures preserved in ice bodies. *Icarus*, *197*(2), 458–469. https://doi.org/10.1016/j.icarus.2008.05.017

Landis, M. E., Byrne, S., Schorghofer, N., Schmidt, B. E., Hayne, P. O., Castillo-Rogez, J., et al. (2017). Conditions for Sublimating Water Ice to Supply Ceres' Exosphere. *J. Geophys. Res. Planets, 122,* 1984-1995. https://doi.org/10.1002/2017JE00533

Landis, M. E., Byrne, S., Combe, J.-Ph., Marchi, S., Castillo-Rogez, J., Sizemore, H. G., Schörghofer, N., Prettyman, T. H., Hayne, P. O., Raymond, C. A., & Russell, C. T. (2019). Water Vapor Contribution to Ceres' Exosphere From Observed Surface Ice and Postulated Ice-Exposing Impacts. *Journal of Geophysical Research: Planets*, *124*(1), 61–75. https://doi.org/10.1029/2018JE005780

Landis, M. E., Prettyman, T. H., Castillo-Rogez, J., & Yamashita, N. (2025). Role of Natron in Delaying the Retreat of Buried Ice Tables on Ceres. *The Planetary Science Journal*, *6*(10), 230. https://doi.org/10.3847/PSJ/ae0333

Liao, S., Hu, J., & Zhang, Y. (2024). Mechanism of hydraulic fracture vertical propagation in deep shale formation based on elastic–plastic model. *Engineering Fracture Mechanics*, *295*, 109806. https://doi.org/10.1016/j.engfracmech.2023.109806

Liestøl, O. (1996). Open-system pingos in Spitsbergen. *Norsk Geografisk Tidsskrift - Norwegian Journal of Geography*, *50*(1), 81–84. https://doi.org/10.1080/00291959608552355

Lei, P., Young, M. W., Seymour, J. D., Stillman, D. E., Primm, K., Sizemore, H. G., Rempel, A. W., & Codd, S. L. (2022). NMR Characterization of unfrozen brine vein distribution and structure in model packed beds. *Cold Regions Science and Technology*, *199*, 103572. https://doi.org/10.1016/j.coldregions.2022.103572

Mackay, J. (1979). Pingos of the Tuktoyaktuk Peninsula Area, Northwest Territories. *Géographie Physique et Quaternaire*, *33*(1), 3–61. https://doi.org/10.7202/1000322ar

Mackay, J. (1998). Pingo Growth and collapse, Tuktoyaktuk Peninsula Area, Western Arctic Coast, Canada: A long-term field study. *Géographie Physique et Quaternaire*, *52*(3), 271–323. https://doi.org/10.7202/004847ar

Mackay, J. R. (1986). Growth of Ibyuk Pingo, Western Arctic Coast, Canada, and some implications for environmental reconstructions. *Quaternary Research*, *26*(1), 68–80. https://doi.org/10.1016/0033-5894(86)90084-0

Marra, W. A., Hauber, E., McLelland, S. J., Murphy, B. J., Parsons, D. R., Conway, S. J., Roda, M., Govers, R., & Kleinhans, M. G. (2014). Pressurized groundwater outflow experiments and numerical modeling for outflow channels on Mars. *Journal of Geophysical Research: Planets*, *119*(12), 2668–2693. https://doi.org/10.1002/2014JE004701

McCord, T. B., & Castillo-Rogez, J. C. (2018). Ceres's internal evolution: The view after Dawn. *Meteoritics & Planetary Science*, *53*(9), 1778–1792. https://doi.org/10.1111/maps.13135

McDonald, M. G., & Harbaugh, A. W. (1988). *A Modular Three-dimensional Finite-difference Ground-water Flow Model*. U.S. Geological Survey.

Mellon, M. T. (1997). Small-scale polygonal features on Mars: Seasonal thermal contraction cracks in permafrost. *Journal of Geophysical Research: Planets*, *102*(E11), 25617–25628. https://doi.org/10.1029/97JE02582

Mellon, M. T., & Phillips, R. J. (2001). Recent gullies on Mars and the source of liquid water. *Journal of Geophysical Research: Planets*, *106*(E10), 23165–23179. https://doi.org/10.1029/2000JE001424

Mellon, M. T., Malin, M. C., Arvidson, R. E., Searls, M. L., Sizemore, H. G., Heet, T. L., Lemmon, M. T., Keller, H. U., & Marshall, J. (2009). The periglacial landscape at the Phoenix landing site. *Journal of Geophysical Research: Planets*, *114*(E1). https://doi.org/10.1029/2009JE003418

Mottaghy, D., & Rath, V. (2006). Latent heat effects in subsurface heat transport modelling and their impact on palaeotemperature reconstructions. *Geophysical Journal International*, *164*(1), 236–245. https://doi.org/10.1111/j.1365-246X.2005.02843.x

Mousis, O., Chassefière, E., Holm, N. G., Bouquet, A., Waite, J. H., Geppert, W. D., Picaud, S., Aikawa, Y., Ali-Dib, M., Charlou, J.-L., & Rousselot, P. (2015). Methane Clathrates in the Solar System. *Astrobiology*, *15*(4), 308–326. https://doi.org/10.1089/ast.2014.1189

Müller, F. (1962). Analysis of Some Stratigraphic Observations and Radiocarbon Dates from Two Pingos in the Mackenzie Delta Area, N. W. T. *Arctic*, *15*(4), 279–288.

Neesemann, A., Gasselt, S. van, Jaumann, R., Castillo-Rogez, J. C., Raymond, C. A., Walter, S. H. G., & Postberg, F. (2025). Improved Stereophotogrammetric and Multi-View Shape-from-Shading DTMs of Occator Crater and Its Interior Cryovolcanism-Related Bright Spots. *Remote Sensing*, *17*(3), 437. https://doi.org/10.3390/rs17030437

Neuzil, C. E. (2019.) Permeability of Clays and Shales. *Annual Review Earth and Planetary Sciences*. 47:247-273. https://doi.org/10.1146/annurev-earth-053018-060437

Neveu, M., & Desch, S. J. (2015). Geochemistry, thermal evolution, and cryovolcanism on Ceres with a muddy ice mantle. *Geophysical Research Letters*, *42*(23), 10,197-10,206. https://doi.org/10.1002/2015GL066375

Ostroumov, V. E., & Siegert, C. (1996). Exobiological aspects of mass transfer in microzones of permafrost deposits. *Advances in Space Research, Proceedings of the F3.1, F3.4, F2.4 and F3.8 Symposia of COSPAR Scientific Commission F*, *18*(12), 79–86. https://doi.org/10.1016/0273-1177(96)00002-6

Palmer, E. M., Heggy, E., Letertre, T., & Sabouroux, P. (2021). Exploring Ceres's Unusual Regolith Porosity and Its Implications for Volatile Retention. *The Planetary Science Journal*, *2*(5), 182. https://doi.org/10.3847/PSJ/ac0b3e

Pamerleau, I. F., Sori, M. M., & Scully, J. E. C. (2024). An ancient and impure frozen ocean on Ceres implied by its ice-rich crust. *Nature Astronomy*, *8*(11), 1373–1379. https://doi.org/10.1038/s41550-024-02350-4

Park, R. S., Konopliv, A. S., Ermakov, A. I., Castillo-Rogez, J. C., Fu, R. R., Hughson, K. H. G., Prettyman, T. H., Raymond, C. A., Scully, J. E. C., Sizemore, H. G., Sori, M. M., Vaughan, A. T., Mitri, G., Schmidt, B. E., & Russell, C. T. (2020). Evidence of non-uniform crust of Ceres from Dawn's high-resolution gravity data. *Nature Astronomy*, *4*(8), 748–755. https://doi.org/10.1038/s41550-020-1019-1

Prettyman, T. H., Yamashita, N., Landis, M. E., Castillo-Rogez, J. C., Schörghofer, N., Pieters, C. M., Sizemore, H. G., Hiesinger, H., Marchi, S., McSween, H. Y., Park, R. S., Toplis, M. J., Raymond, C. A., & Russell, C. T. (2021). Replenishment of Near-Surface Water Ice by Impacts Into Ceres' Volatile-Rich Crust: Observations by Dawn's Gamma Ray and Neutron Detector. *Geophysical Research Letters*, *48*(15), e2021GL094223. https://doi.org/10.1029/2021GL094223

Prettyman, T. H., Yamashita, N., Toplis, M. J., McSween, H. Y., Schörghofer, N., Marchi, S., Feldman, W. C., Castillo-Rogez, J., Forni, O., Lawrence, D. J., Ammannito, E., Ehlmann, B. L., Sizemore, H. G., Joy, S. P., Polanskey, C. A., Rayman, M. D., Raymond, C. A., & Russell, C. T. (2017). Extensive water ice within Ceres' aqueously altered regolith: Evidence from nuclear spectroscopy. *Science*, *355*(6320), 55–59. https://doi.org/10.1126/science.aah6765

Quick, L. C., Buczkowski, D. L., Ruesch, O., Scully, J. E. C., Castillo-Rogez, J., Raymond, C. A., Schenk, P. M., Sizemore, H. G., & Sykes, M. V. (2019). A Possible Brine Reservoir Beneath Occator Crater: Thermal and Compositional Evolution and Formation of the Cerealia Dome and Vinalia Faculae. *Icarus, Occator Crater on Ceres*, *320*, 119–135. https://doi.org/10.1016/j.icarus.2018.07.016

Raponi, A., De Sanctis, M. C., Carrozzo, F. G., Ciarniello, M., Castillo-Rogez, J. C., Ammannito, E., Frigeri, A., Longobardo, A., Palomba, E., Tosi, F., Zambon, F., Raymond, C. A., & Russell, C. T. (2019). Mineralogy of Occator crater on Ceres and insight into its evolution from the properties of carbonates, phyllosilicates, and chlorides. *Icarus, Occator Crater on Ceres*, *320*, 83–96. https://doi.org/10.1016/j.icarus.2018.02.001

Raymond, C. A., Ermakov, A. I., Castillo-Rogez, J. C., Marchi, S., Johnson, B. C., Hesse, M. A., Scully, J. E. C., Buczkowski, D. L., Sizemore, H. G., Schenk, P. M., Nathues, A., Park, R. S., Prettyman, T. H., Quick, L. C., Keane, J. T., Rayman, M. D., & Russell, C. T. (2020a). Impact-driven mobilization of deep crustal brines on dwarf planet Ceres. *Nature Astronomy*, *4*(8), 741–747. https://doi.org/10.1038/s41550-020-1168-2

Rempel, A. W. (2007). Formation of ice lenses and frost heave. *Journal of Geophysical Research: Earth Surface*, *112*(F2). https://doi.org/10.1029/2006JF000525

Rempel, A. W. (2008), A theory for ice-till interactions and sediment entrainment beneath glaciers, *J. Geophys. Res., 113,* F01013, doi:10.1029/2007JF000870.

Rivkin, A. S., Li, J.-Y., Milliken, R. E., Lim, L. F., Lovell, A. J., Schmidt, B. E., McFadden, L. A., & Cohen, B. A. (2011). The surface composition of Ceres. *Space Science Reviews*, **163**(1), 95–116, https://doi.org/10.1007/978-1-4614-4903-4_7

Roatsch, Th., Kersten, E., Matz, K.-D., Preusker, F., Scholten, F., Jaumann, R., Raymond, C. A., & Russell, C. T. (2017). High-resolution Ceres Low Altitude Mapping Orbit Atlas derived from Dawn Framing Camera images. *Planetary and Space Science*, *140*, 74–79. https://doi.org/10.1016/j.pss.2017.04.008

Roatsch, Th., Kersten, E., Matz, K.-D., Jaumann, R., Raymond, C. A. (2018). Highest resolution mosaic of Ceres derived from Dawn Framing Camera. *American Geophysical Union Meeting,* #P33D-3869.

Ruesch, O., Quick, L. C., Landis, M. E., Sori, M. M., Čadek, O., Brož, P., Otto, K. A., Bland, M. T., Byrne, S., Castillo-Rogez, J. C., Hiesinger, H., Jaumann, R., Krohn, K., McFadden, L. A., Nathues, A., Neesemann, A., Preusker, F., Roatsch, T., Schenk, P. M., … Russell, C. T. (2019). Bright carbonate surfaces on Ceres as remnants of salt-rich water fountains. *Icarus, Occator Crater on Ceres*, *320*, 39–48. https://doi.org/10.1016/j.icarus.2018.01.022

Scheidegger, J. M., V. F.Bense, and S. E.Grasby (2012), Transient nature of Arctic spring systems driven by subglacial meltwater, *Geophys. Res. Lett., 39,* L12405, doi:10.1029/2012GL051445.

Schenk, P., Scully, J., Buczkowski, D., Sizemore, H., Schmidt, B., Pieters, C., Neesemann, A., O'Brien, D., Marchi, S., Williams, D., Nathues, A., De Sanctis, M., Tosi, F., Russell, C. T., Castillo-Rogez, J., & Raymond, C. (2020). Impact heat driven volatile redistribution at Occator crater on Ceres as a comparative planetary process. *Nature Communications*, *11*(1), 3679. https://doi.org/10.1038/s41467-020-17184-7

Schenk, P., Sizemore, H., Schmidt, B., Castillo-Rogez, J., De Sanctis, M., Bowling, T., Scully, J., Buczkowski, D., Quick, L., Preusker, F., Park, R., Raymond, C., & Russell, C. (2019). The central pit and dome at Cerealia Facula bright deposit and floor deposits in Occator crater, Ceres: Morphology, comparisons and formation. *Icarus, Occator Crater on Ceres*, *320*, 159–187. https://doi.org/10.1016/j.icarus.2018.08.010

Schmidt, B. E., Hughson, K. H. G., Chilton, H. T., Scully, J. E. C., Platz, T., Nathues, A., Sizemore, H., Bland, M. T., Byrne, S., Marchi, S., O'Brien, D. P., Schorghofer, N., Hiesinger, H., Jaumann, R., Pasckert, J. H., Lawrence, J. D., Buzckowski, D., Castillo-Rogez, J. C., Sykes, M. V., … Raymond, C. A. (2017). Geomorphological evidence for ground ice on dwarf planet Ceres. *Nature Geoscience*, *10*(5), 338–343. https://doi.org/10.1038/ngeo2936

Schmidt, B. E., Sizemore, H. G., Hughson, K. H. G., Duarte, K. D., Romero, V. N., Scully, J. E. C., Schenk, P. M., Buczkowski, D. L., Williams, D. A., Nathues, A., Udell, K., Castillo-Rogez, J. C., Raymond, C. A., & Russell, C. T. (2020). Post-impact cryo-hydrologic formation of small mounds and hills in Ceres's Occator crater. *Nature Geoscience*, *13*(9), 605–610. https://doi.org/10.1038/s41561-020-0581-6

Schorghofer, N. (2008). The Lifetime of Ice on Main Belt Asteroids. *The Astrophysical Journal*, *682*(1), 697. https://doi.org/10.1086/588633

Schorghofer, N. (2009). Buffering of sublimation loss of subsurface ice by percolating snowmelt: A theoretical analysis. *Permafrost and Periglacial Processes*, *20*(3), 309–313. https://doi.org/10.1002/ppp.646

Schorghofer, N. (2016). Predictions of depth-to-ice on asteroids based on an asynchronous model of temperature, impact stirring, and ice loss. *Icarus*, *276*, 88–95. https://doi.org/10.1016/j.icarus.2016.04.037

Schorghofer, N., Mazarico, E., Platz, T., Preusker, F., Schröder, S. E., Raymond, C. A., & Russell, C. T. (2016). The permanently shadowed regions of dwarf planet Ceres. *Geophysical Research Letters*, *43*(13), 6783–6789. https://doi.org/10.1002/2016GL069368

Scully, J. E. C., Buczkowski, D. L., Raymond, C. A., Bowling, T., Williams, D. A., Neesemann, A., Schenk, P. M., Castillo-Rogez, J. C., & Russell, C. T. (2019). Ceres' Occator crater and its faculae explored through geologic mapping. *Icarus, Occator Crater on Ceres*, *320*, 7–23. https://doi.org/10.1016/j.icarus.2018.04.014

Scully, J. E. C., Schenk, P. M., Castillo-Rogez, J. C., Buczkowski, D. L., Williams, D. A., Pasckert, J. H., Duarte, K. D., Romero, V. N., Quick, L. C., Sori, M. M., Landis, M. E., Raymond, C. A., Neesemann, A., Schmidt, B. E., Sizemore, H. G., & Russell, C. T. (2020). The varied sources of faculae-forming brines in Ceres' Occator crater emplaced via hydrothermal brine effusion. *Nature Communications*, *11*(1), 3680. https://doi.org/10.1038/s41467-020-15973-8

Singhal †, B. B. S., & Gupta, R. P. (2010). *Applied Hydrogeology of Fractured Rocks: Second Edition*. Springer Science & Business Media.

Sizemore, H. G., & Mellon, M. T. (2008). Laboratory characterization of the structural properties controlling dynamical gas transport in Mars-analog soils. *Icarus*, *197*(2), 606–620. https://doi.org/10.1016/j.icarus.2008.05.013

Sizemore, H. G., Mellon, M. T., Rempel, A. W., McKay, C. P., Heldmann, J. L., Stoker, C. R., & Perry, M. R. (2026). Liquid Vein networks as habitats in ice-cemented ground on Earth and Mars: Effects of soil geometry and salts. *Icarus*, *445*, 116828. https://doi.org/10.1016/j.icarus.2025.116828

Sizemore, H. G., Platz, T., Schorghofer, N., Prettyman, T. H., Sanctis, M. C. D., Crown, D. A., Schmedemann, N., Neesemann, A., Kneissl, T., Marchi, S., Schenk, P. M., Bland, M. T., Schmidt, B. E., Hughson, K. H. G., Tosi, F., Zambon, F., Mest, S. C., Yingst, R. A., Williams, D. A., … Raymond, C. A. (2017). Pitted terrains on (1) Ceres and implications for shallow subsurface volatile distribution*, Geophys. Res. Lett., 44,* 6570–6578, doi:10.1002/2017GL073970.

Sizemore, H. G., Schmidt, B. E., Buczkowski, D. A., Sori, M. M., Castillo-Rogez, J. C., Berman, D. C., Ahrens, C., Chilton, H. T., Hughson, K. H. G., Duarte, K., Otto, K. A., Bland, M. T., Neesemann, A., Scully, J. E. C., Crown, D. A., Mest, S. C., Williams, D. A., Platz, T., Schenk, P., … Raymond, C. A. (2019). A Global Inventory of Ice-Related Morphological Features on Dwarf Planet Ceres: Implications for the Evolution and Current State of the Cryosphere. *Journal of Geophysical Research: Planets*, *124*(7), 1650–1689. https://doi.org/10.1029/2018JE005699

Sizemore, H. G., Zent, A. P., & Rempel, A. W. (2015). Initiation and growth of martian ice lenses. *Icarus, Dynamic Mars*, *251*, 191–210. https://doi.org/10.1016/j.icarus.2014.04.013

Sori, M. M., Sizemore, H. G., Byrne, S., Bramson, A. M., Bland, M. T., Stein, N. T., & Russell, C. T. (2018). Cryovolcanic rates on Ceres revealed by topography. *Nature Astronomy*, *2*(12), 946–950. https://doi.org/10.1038/s41550-018-0574-1

Stein, N. T., Ehlmann, B. L., Palomba, E., De Sanctis, M. C., Nathues, A., Hiesinger, H., Ammannito, E., Raymond, C. A., Jaumann, R., Longobardo, A., & Russell, C. T. (2019). The formation and evolution of bright spots on Ceres. *Icarus, Occator Crater on Ceres*, *320*, 188–201. https://doi.org/10.1016/j.icarus.2017.10.014

Titus, T. N. (2015). Ceres: Predictions for near-surface water ice stability and implications for plume generating processes. *Geophysical Research Letters*, *42*(7), 2130–2136. https://doi.org/10.1002/2015GL063240

Tornabene, L. L., Osinski, G. R., McEwen, A. S., Boyce, J. M., Bray, V. J., Caudill, C. M., Grant, J. A., Hamilton, C. W., Mattson, S., & Mouginis-Mark, P. J. (2012). Widespread crater-related pitted materials on Mars: Further evidence for the role of target volatiles during the impact process. *Icarus*, *220*(2), 348–368. https://doi.org/10.1016/j.icarus.2012.05.022

Williams, P. J., & Smith, M. W. (1989). *The Frozen Earth: Fundamentals of Geocryology*. Cambridge University Press. https://doi.org/10.1017/CBO9780511564437

Xu, X., Wang, B., Fan, C., & Zhang, W. (2020). Strength and deformation characteristics of silty clay under frozen and unfrozen states. *Cold Regions Science and Technology*, *172*, 102982. https://doi.org/10.1016/j.coldregions.2019.102982

Yoshikawa, K., & Harada, K. (1995). Observations on nearshore pingo growth, Adventdalen, Spitsbergen. *Permafrost and Periglacial Processes*, *6*(4), 361–372. https://doi.org/10.1002/ppp.3430060407

You, Z., Ma, Y., Wang, Z., & Ma, J. (2021). Tensile strength variation of a silty clay under different temperature and moisture conditions. *Cold Regions Science and Technology*, *189*, 103314. https://doi.org/10.1016/j.coldregions.2021.103314

Zolotov, M. Yu. (2017). Aqueous origins of bright salt deposits on Ceres. *Icarus*, *296*, 289–304. https://doi.org/10.1016/j.icarus.2017.06.018